# A new Time-decay Radiomics Integrated Network (TRINet) for short-term breast cancer risk prediction

Hong Hui Yeoh[1], Fredrik Strand[3,4], Raphaël Phan[5], Kartini Rahmat[6], Maxine Tan[1,2,*]

[1] Department of Electrical and Robotics Engineering, School of Engineering, Monash University Malaysia, Bandar Sunway 47500, Malaysia.

[2] School of Electrical and Computer Engineering, The University of Oklahoma, Norman, OK 73019, USA.

[3] Department of Oncology and Pathology, Karolinska Institute, Solna, Sweden.

[4] Breast Radiology, Karolinska University Hospital, Solna, Sweden.

[5] School of Information Technology, Monash University Malaysia, Bandar Sunway 47500, Malaysia.

[6] Department of Biomedical Imaging and University of Malaya Research Imaging Centre, Faculty of Medicine, University of Malaya, 50603 Kuala Lumpur, Malaysia.

[*] Corresponding author. Email address: Maxine.Tan@monash.edu

## Abstract

To facilitate early detection of breast cancer, there is a need to develop short-term risk prediction schemes that can prescribe personalized/individualized screening mammography regimens for women. In this study, we propose a new deep learning architecture called TRINet that implements time-decay attention to focus on recent mammographic screenings, as current models do not account for the relevance of newer images. We integrate radiomic features with an Attention-based Multiple Instance Learning (AMIL) framework to weigh and combine multiple views for better risk estimation. In addition, we introduce a continual learning approach with a new label assignment strategy based on bilateral asymmetry to make the model more adaptable to asymmetrical cancer indicators. Finally, we add a time-embedded additive hazard layer to perform dynamic, multi-year risk forecasting based on individualized screening intervals. We used two public datasets, namely 8,528 patients from the American EMBED dataset and 8,723 patients from the Swedish CSAW dataset in our experiments. Evaluation results on the EMBED test set show that our approach significantly outperforms state-of-the-art models, achieving AUC scores of 0.851, 0.811, 0.796, 0.793, and 0.789 across 1-, 2-, to 5-year intervals, respectively. Our results underscore the importance of integrating temporal attention, radiomic features, time embeddings, bilateral asymmetry, and continual learning strategies, providing a more adaptive and precise tool for short-term breast cancer risk prediction.

# 1. Introduction

There is a clear need to develop effective breast cancer screening protocols and individualized/personalized risk assessment models (Dadsetan et al., 2022; Lee et al., 2023b; Yala et al., 2021; Yeoh et al., 2023). However, current screening protocols are generic, based on fixed schedules and broad risk categories (Ren et al., 2022). This can lead to over-screening, causing harm in the form of false positives, unnecessary anxiety, and overtreatment for those who have low cancer risk (Bond et al., 2013; Habib et al., 2021; Salz et al., 2010). Conversely, these fixed screening protocols could also lead to under-screening and missing early cancer signs in high-risk women, leading to cancers being detected too late.

Image based risk assessment modelling is underutilized in current clinical practice although recent artificial intelligence (AI) studies report significantly-improved performance using mammography data (Arasu et al., 2023). Recent deep learning advances allow us to build risk assessment models that analyse mammograms directly and show significant improvements over conventional/statistical models (Dadsetan et al., 2022; Lee et al., 2023b; Lotter et al., 2021; McKinney et al., 2020; Yala et al., 2021). These models use convolutional neural networks (CNNs) to learn patterns in images and outperform traditional models that use manually selected/handcrafted features (Gastounioti et al., 2016; Tan et al., 2019; Tan et al., 2016).

While these models demonstrate improvements over previous conventional methods, one of the biggest limitations of current models is that they are designed to analyse a single static mammogram, ignoring the temporal progression of mammographic changes across multiple screenings (Donnelly et al., 2024; Lotter et al., 2021; McKinney et al., 2020; Tan et al., 2019; Yala et al., 2021). As cancer is a dynamic process and structural changes in breast tissue progress gradually over time, sequential mammograms that capture these changes contain critical information for identifying emerging cancer risks. Without incorporating temporal data, current models may miss the subtle yet critical signs of risk progression, thus limiting their predictive ability. Only a few studies (Dadsetan et al., 2022; Lee et al., 2023b; Yeoh et al., 2023) in the literature have included previous mammographic screenings in training their model.

Another drawback of current risk models is that although both parenchymal texture/radiomic features and deep learning features have proved beneficial for risk assessment, only a few studies combine both radiomic and deep learning features in their models (Yeoh et al., 2023). The simplest way to combine both feature groups is to concatenate them in the fully-connected (FC) layer (Yeoh et al., 2023). However, examining more advanced, optimal methods (Bahdanau et al., 2014; Ilse et al., 2018) is required to combine these features to maximally leverage both feature groups' strengths and to synthesize new combinations of features for better model performance. Additionally, models that only utilize either the craniocaudal (CC) or mediolateral oblique (MLO) view (Arefan et al., 2019; Carneiro et al., 2015) to form a final risk prediction score are outperformed by models that incorporate both views. Conversely, we propose a new model that combines sequential radiomic and deep learning based image features from all four images of both views to generate a final comprehensive risk score for individual patients.

Furthermore, although implementing a personalized/individualized screening regime is highly recommended to avoid over-screening or under-screening of women, to date, none of the risk models incorporate this information into their models to better stratify women in a short-term risk prediction scheme. We present a new time-interval embedding method for better risk prediction/classification in the feature space that can forecast the risk of cancer occurring in 6 months to 5 years' time. Incorporation of this new time-embedding method can help stratify women into personalized screening regimens to avoid over-screening in women who have a low short-term cancer risk and can thus come back later (say in 3 years' time) for their next screening and avoid under-screening women who have suspicious signs and thus have high short-term cancer risk and should come back earlier (say in 6 months' time) for their next screening.

Another issue that is present not just in risk prediction schemes but in AI schemes in the medical field is that the full capacity of current methods might be limited by their inability to continually learn (Liu et al., 2024; Wang et al., 2023). Current risk prediction models are only trained on a single dataset of a single demographic group/population, e.g., a Swedish (Yeoh et al., 2023), American (Yala et al., 2021), or Asian (Tan et al., 2019) population. If a new dataset becomes available, it would be most unfortunate if the trained model suffers from catastrophic forgetting and is unable to increase its knowledge through a continual learning framework/method, given how difficult/challenging it is to obtain ethics clearances, etc., for any given dataset. To resolve this issue, we present a new method that enables a model to continually learn and increase its knowledge using a new dataset of a different population without catastrophically forgetting its past/previous knowledge of the primary dataset.

Additionally, bilateral asymmetry – differences in parenchymal tissue between left and right breasts – is a well-known risk factor as cancers typically develop in only one breast over time (Tan et al., 2016; Zheng et al., 2012). However, many risk models with the exception of a few (Donnelly et al., 2024; Tan et al., 2016; Yeoh et al., 2023), do not incorporate this measurement as indicators in their models. In this study, we propose a new approach to measuring bilateral asymmetry and incorporate it with our continual learning method to enhance our model's performance. The results show that using our new approach, the model can continually learn on a new dataset and improve its performance without catastrophic forgetting of the previous dataset.

Thus, to overcome all these limitations, we present a Time-decay Radiomics Integrated Network (TRINet), which is a new deep learning cancer risk model with the following contributions:

- We present a new Time-Decay (TD) attention mechanism for both Non-Local (Wang et al., 2018) and Fastformer (Wu et al., 2021) blocks. Although TD attention improves both Non-Local and Fastformer performances, its application on the Fastformer block yields better overall performance.

- We introduce a new Radiomics and Deep learning feature based Multiple Instance Learning (RADMIL) method to effectively integrate and combine deep learning and radiomics features. RADMIL effectively integrates features from CC and MLO views of both left and right breasts in an interpretable and effective manner using attention mechanisms.

- We propose a new cancer forecast network that incorporates time-interval embeddings for short-term risk prediction in six-month intervals (i.e., 6 months to 5 years). In our new method, a time-embedded additive hazard layer is presented for short-term risk prediction. The time embedding provides context/labels to the risk model to forecast individualized screening intervals better.

- We present a new self-training method based on continual learning called Reinforced Self-Training (ReST) with Continual Learning (ReST$^{CL}$). ReST$^{CL}$ enables our model to continually learn new knowledge on new/secondary datasets without catastrophically forgetting the previous/primary dataset. To stratify/identify useful samples for finetuning from the secondary dataset, bilateral asymmetry features are computed, and the model is trained iteratively. Results show that new knowledge is gained on the secondary dataset of a different population/demographic without catastrophically forgetting previous knowledge of the primary dataset.

## 2. Related work

### 2.1 Short-term breast cancer risk prediction models

The development of breast cancer risk prediction for mammographic images has come a long way. The earlier methods were based on radiomics features including parenchymal texture (Anandarajah et al., 2021; Tan et al., 2016) and mammographic density (Anandarajah et al., 2021; Keller et al., 2012) based features. More recently, deep learning (Dadsetan et al., 2022; Lee et al., 2023b; Yala et al., 2021; Yeoh et al., 2023; Zhu et al., 2021) features have been incorporated into newer models. However, the vast majority of existing methods do not incorporate temporal/longitudinal information in the form of prior screening mammograms with the exception of a few methods (Dadsetan et al., 2022; Lee et al., 2023b; Yeoh et al., 2023) that we will summarize in this subsection.

First, Mirai (Yala et al., 2021) focuses on robustness and generalizability in handling missing clinical risk factor information and consistent predictions across different mammography machines. However, Mirai has a key limitation of using single time-point mammograms for risk prediction. By ignoring longitudinal information, Mirai cannot capture temporal changes in breast tissue that can be critical to cancer risk progression, which is an issue that was acknowledged by the authors in their paper(s). While Mirai obtains strong population-level performance on test datasets from different institutions, it does not utilize sequential mammograms to provide more personalized/dynamic risk assessments.

LRP-NET (Dadsetan et al., 2022) models spatiotemporal changes in breast tissue across multiple sequential mammograms, to capture longitudinal changes/information for any given patient. This is achieved through capturing image features using a CNN encoder and modelling temporal changes with a Gated Recurrent Unit (GRU). The highest Area under the Receiver Operating Characteristic Curve (AUC) result of 67% is reported for information obtained from four prior screening mammograms. However, LRP-NET does not incorporate radiomics features in its model, which can provide additional information to deep learning features. It also does not address short-term risk forecasting (e.g., 1-year or 2-year AUC) for the implementation of personalized/individualized screening regiments.

Another model is PRIME+ (Lee et al., 2023b) whose network architecture is similar to Mirai's and implements a CNN encoder with a transformer decoder. PRIME+ includes a cross-attention mechanism in its transformer block, whereby queries are obtained from the current exam, and keys and values from the prior exams. This allows PRIME+ to capture temporal changes in breast tissue, e.g. density changes, and outperforms single time-point models. However, like Mirai and LRP-NET, PRIME+ does not incorporate radiomics features nor continual learning methods for better performance.

## 2.2 Time-Decay (TD) attention

Time-sensitive time-decay attention plays an important role across diverse domains, addressing the need for deep learning models to effectively adapt to temporal variations. Notably, in the field of Natural Language Processing (NLP), Receptance Weighted Key Value (RWKV) (Peng et al., 2023) presents an innovative approach to reconcile the trade-offs between computational efficiency and model performance in sequence processing. The authors introduce a novel model architecture, which combines the strengths of transformers and recurrent neural networks (RNNs). RWKV draws inspiration from the Attention Free Transformer (AFT) (Zhai et al., 2021) and introduces a channel-wise exponential time decay vector, which is constrained to be non-negative to ensure proper decay characteristics. This is multiplied by the relative position and traced backward from the current time, facilitating the attention mechanism to focus more consciously on recent inputs. In another similar architecture, the Retentive Network (RETNET) (Sun et al., 2023) instead employs exponential decay within the causal mask, which combines with the query-key matrix through the dot product. RETNET introduces a retention mechanism for sequence modelling, supporting various computation paradigms such as parallel, recurrent, and chunk-wise recurrent representations.

A notable contribution in the medical field is GLIM-Net (Hu et al., 2023), a Chronic Glaucoma Forecast Transformer, which introduces time positional encoding to learn temporal information from sequential fundus images. GLIM-Net achieves time-sensitive attention through a time-related matrix $T$, combined with the attention matrix through the Hadamard product to better enable the self-attention mechanism to handle irregularly sampled data. Another paper (Li et al., 2022) introduced two extensions to the standard Vision Transformer (ViT) (Dosovitskiy et al., 2021) for the task of lung cancer diagnosis from longitudinal computed tomography. The authors introduced continuous-time vector embeddings by constructing a relative time distance vector to incorporate linear time information into the transformer, allowing the model to capture temporal dependencies between the images. Additionally, they proposed a time-aware ViT that learns through a flipped sigmoid function to scale self-attention weights at each head, addressing the challenge of decreasing information over irregularly sampled time intervals.

## 2.3 Radiomic features for breast cancer risk prediction

Radiomics (Lambin et al., 2012), the extraction and quantification of imaging biomarkers, is important to advance cancer risk prediction but is seldom leveraged for this purpose. By capturing quantitative information such as texture, shape, and intensity patterns from images, radiomics can uncover features that are not visible to the human eye.

Radiomics has been applied to predict breast cancer risk factors including molecular subtypes and recurrence risks. Ma et al. developed radiomics-based mammographic classifiers to differentiate between triple-negative (TN) and non-TN cancers as well as HER2-enriched and luminal breast cancers with an AUC of up to 86.5% (Ma et al., 2019). Kontos et al. applied radiomics classifiers to identify mammographic parenchymal complexity phenotypes associated with cancer risk and reported an AUC of 84% in their study (Kontos et al., 2019).

Deep-LIBRA (Maghsoudi et al., 2021) is an important artificial intelligence (AI) method using radiomics for breast cancer risk assessment. This method combines CNNs for breast segmentation with a radiomic algorithm to differentiate dense and non-dense breast tissue. Deep-LIBRA quantifies breast percent density (PD), a well-established risk factor [15-18] for breast cancer. The study results show that Deep-

LIBRA outperforms several state-of-the-art breast density assessment methods in case-control discrimination with an AUC of 61.2%. While these results demonstrate the potential of combining deep learning and radiomics for risk prediction, the approach only focuses on breast density as a single biomarker and does not incorporate prior mammographic images in its risk assessment.

While all these methods present important contributions to cancer risk assessment models, there is a need to develop an integrated and comprehensive model that combines all unique and individual characteristics into an all-encompassing model. In the subsequent sections, we introduce our novel TRINet model, whose overall architecture is shown in Fig. 1, which implements TD attention, effectively integrates radiomic features and time-interval embeddings, and continually learns on new datasets for more accurate risk prediction.

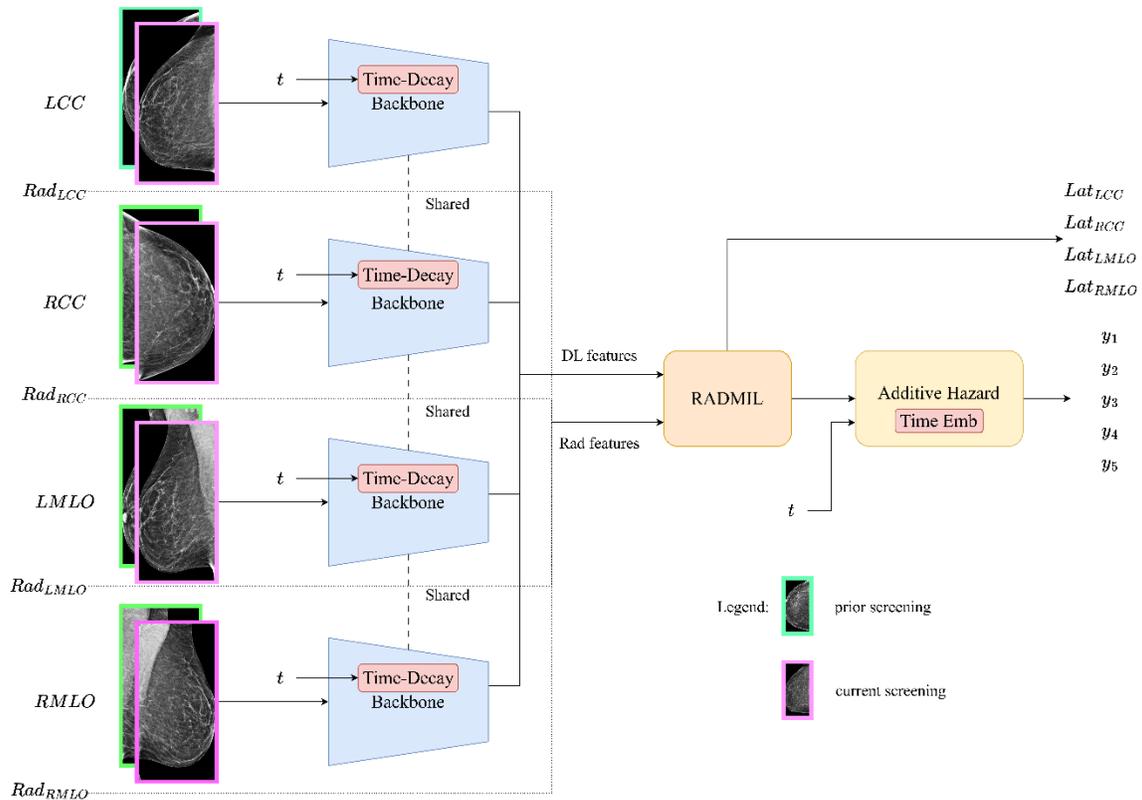

Fig. 1: The proposed TRINet model uses a shared backbone encoder across all 4 images. The encoder incorporates a new time-decay attention that can capture changes in breast tissue structure over time while focusing on more recent screenings. The handcrafted radiomic features are the same as in (Yeoh et al., 2023), and are included into the model through our new Radiomics and Deep learning based Multiple Instance Learning (RADMIL) architecture.

## 3. Materials and Methods

## 3.1 Datasets

### 3.1.1 EMBED dataset

The Emory Breast Imaging Dataset (EMBED) (Jeong et al., 2023) contains over 3.4 million images from 116,000 women including a balanced mix of African American and white patients. EMBED addresses the diversity and granularity gaps in breast imaging datasets. This dataset includes 2D and DBT images along with lesion-level annotations and longitudinal follow-up data. Demographic data (including age, race, insurance status) and clinical data are included as well as lesion ground-truth pathology data including outcomes from biopsy and surgery, categorized into 7 severity levels from benign to invasive cancer.

By linking imaging features to pathology-confirmed outcomes, EMBED's sequential data allows to train and validate machine learning models that account for changes in mammographic findings over time. This is crucial for risk prediction as models can be trained to recognize subtle changes that may indicate increasing risk over time and potentially allow for early intervention. The dataset's racial diversity enables the development of risk models that work well across different demographics, addressing known biases in cancer screening and reducing disparities in early detection and outcomes. The dataset used in this study is the 20% portion of the dataset that is "open" for research access, thus facilitating performance comparisons with other methods in the literature.

describes our process/procedure of filtering this dataset for the purpose of this study.

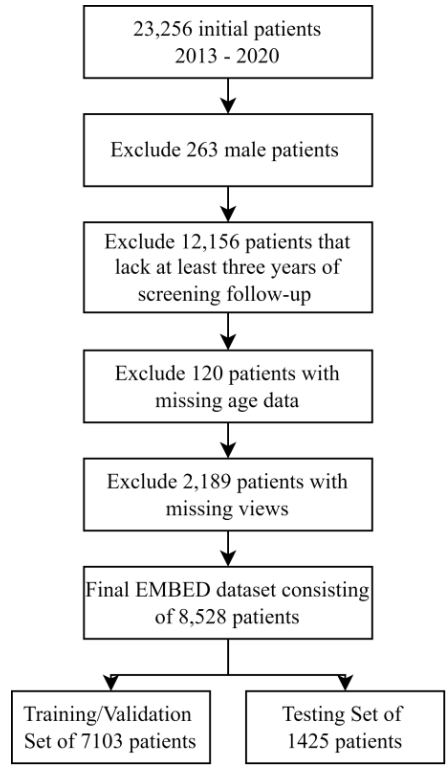

Fig. 2: Flowchart for exam selection of the EMBED dataset used in this study.

EMBED has longitudinal data that tracks the number and frequency of screenings for patients, including those who go on to develop breast cancer. Patients in the dataset have multiple screenings over time, reflecting real world scenarios where repeated imaging is key to early detection and monitoring of disease progression. This sequential imaging data allows researchers to study patterns in mammographic changes leading up to a diagnosis, such as the development or evolution of lesions. This data is particularly valuable for understanding how screening intervals and adherence impact outcomes and enable the optimization of personalized screening protocols. Fig. 3 shows the breakdown of the number of screenings for cancer patients.

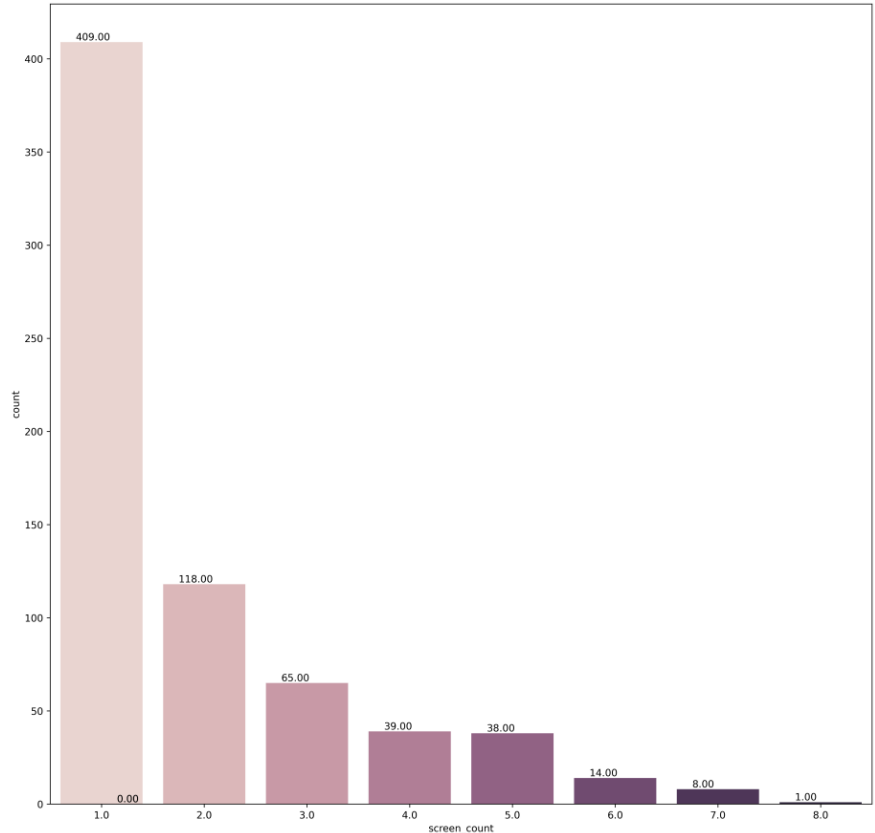

Fig. 3: Breakdown of number of screenings for cancer patients.

### 3.1.2 CSAW dataset

In this study, we utilize the same Cohort of Screen-Aged Women (CSAW) (Dembrower et al., 2020) dataset as in our previous work (Yeoh et al., 2023). This dataset, previously described in detail in (Yeoh et al., 2023), comprises of a population-based cohort of women aged 40–74 years old from the Stockholm region, Sweden, with mammograms collected between 2008 and 2016. For consistency and reproducibility, we applied the same data processing/filtering strategy used in (Yeoh et al., 2023), maintaining the control and case partitioning, along with the categorization of clinical risk factors. The proportion of cases and controls was also maintained for the training, validation, and testing sets.

## 3.2 Time-Decay (TD) attention

Traditional attention mechanisms weigh each pixel in spacetime equally, giving equal importance to each pixel's position (Wang et al., 2018). However, in a scenario where the individual image frame progresses through time, the relevance of information changes over time, and an equally weighted attention mechanism may not suffice. To address this issue, we propose to incorporate the concept of time decay into the attention mechanism. Our aim is to leverage existing insights that earlier images carry less relevant information as compared with more recent ones. In line with this, attention mechanisms should selectively pay less attention to historical/prior mammographic features and focus more on the most up-to-date/recent images. This emulates the diagnostic approach of radiologists who, while referencing prior mammograms, place greater emphasis on the most current images in making their final diagnosis (Hayward et al., 2016).

The novelties in this section draw inspiration from the time-sensitive self-attention mechanism in (Hu et al., 2023) proposed for glaucoma forecasting using time interval data, the RWKV (Peng et al., 2023), and Retentive Network (Sun et al., 2023) for large language models. We hereby propose a new time-decaying attention for the Non-Local self-attention (Wang et al., 2018) and Spatial Channel Image Fastformer (SHIFT) blocks (Yeoh et al., 2023). A time-sensitive mechanism was proposed in (Hu et al., 2023) in a Transformer (Vaswani et al., 2017) architecture with Encoder-Decoder blocks; however, Transformers generally require an abundance of images to be trained efficiently, which might not be available in the medical imaging field (Dosovitskiy et al., 2021). Conversely, we propose new time-decay attention mechanisms in CNN architectures that generally require fewer images to train. Another advantage is that in CNN architectures, 3D convolution can be implemented to find useful information in the time dimension, e.g., in a sequence of yearly mammograms. By incorporating time decay into the attention mechanism, we aim to enhance the model's ability to discern temporal patterns and direct its focus appropriately, ultimately improving its performance in tasks where recent information holds more significance than past observations.

### 3.2.1 Time-Decay Non-Local (TD-NL) block

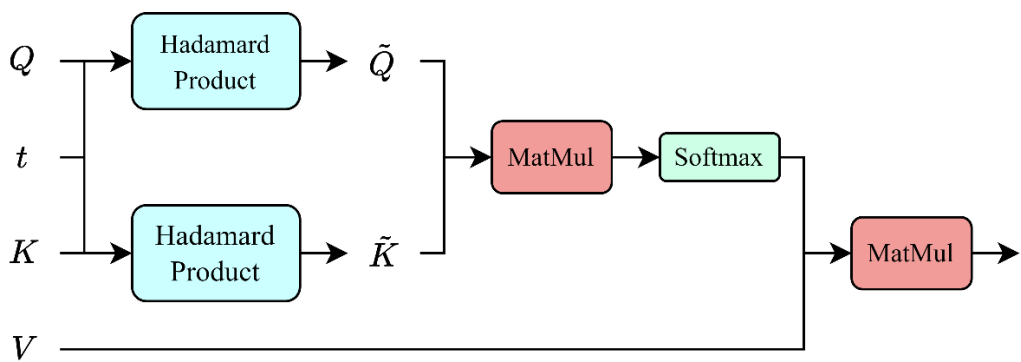

Fig. 4: Time-decay Non-Local (TD-NL) attention block

To implement time-decay attention, our proposed method builds upon the attention mechanisms of Non-Local networks (Wang et al., 2018) and our previously-proposed SHIFT (Yeoh et al., 2023) block. In this subsection, we focus on formulating a new time-decay attention for non-local; we will propose a similar definition for SHIFT in the next subsection. We first linearly transform our input $X \in \mathbb{R}^{C \times n}$, where $C$ represents the channel dimension and $n$ denotes $time \times height \times width$, respectively, into query, key, and value matrices of $Q, K, V \in \mathbb{R}^{\bar{C} \times n}$. The formulation can be expressed as follows:

$$Q = M_Q X, \qquad (1)$$
$$K = M_K X, \qquad (2)$$
$$V = M_V X, \qquad (3)$$

where $M_Q, M_K$ and $M_V$ are $1 \times 1$ convolutions. To incorporate a time interval vector, $t$ as an input into our modified attention block, Hadamard product in the time dimension is used. For Non-Local self-attention, the time-decay attention mechanism can thus be formally represented as follows:

$$NL(Q, K, V, t) = softmax(\tilde{Q}^T \tilde{K})V, \qquad (4)$$
$$\tilde{Q} = Q * t, \qquad (5)$$
$$\tilde{K} = K * t, \qquad (6)$$
$$t = \frac{1}{e^A e^{B \Delta t_{i,n}}}, \qquad (7)$$
$$\Delta t_{i,n} = \frac{\min(\Delta t_{i,n}, T)}{T} \qquad (8)$$

where $A$ and $B$ are fixed parameters and $\Delta t_{i,n}$ is the time interval/difference between images in the sequence, $x_i$ and the most recent/ "current" image in the sequence, $x_n$. As patients generally have follow-up screenings in a year's time, we measure time intervals in months. Threshold $T$ is used to clip the time interval, $\Delta t_{i,n}$. Following that, the resulting value is normalized by dividing it by $T$ to normalize the range of $\Delta t_{i,n}$ to be between 0 and 1. The threshold value $T$ in the formula can be chosen arbitrarily; however, it is recommended to select a value that is not too high, as patients tend to come annually for their screening mammogram as typified in our dataset(s). The negative exponential in Eq. 7 is a monotonically decreasing function, which ensures decaying/less attention is given to "prior" mammograms that are further away from the "current" mammogram in time.

Our formulation and implementation for handling time attention differs from (Hu et al., 2023). First and foremost, we avoided the use of transformers, which typically demand a substantial amount of training samples. Unlike the transformer-based approaches to handle time attention, our method directly influences the query and key matrices through Hadamard product with the time interval vector. This direct integration of temporal information into the query and key matrices eliminates the need for constructing and learning a separate query-key interaction matrix with more trainable parameters, contributing to a more efficient and sample-efficient approach.

### 3.2.2 Time-Decay SHIFT (TD-SHIFT) block

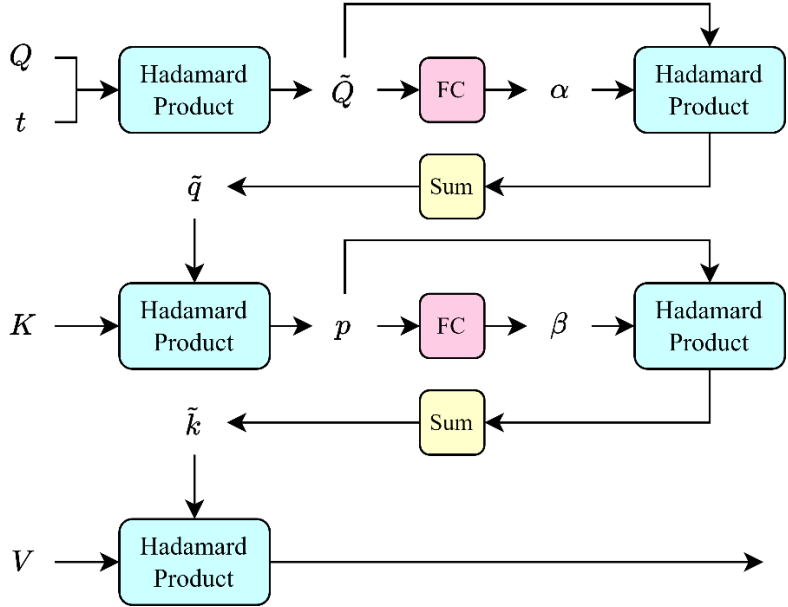

Fig. 5: Time-decay SHIFT (TD-SHIFT) attention block

In this subsection, we extend the time-decay attention to our SHIFT block proposed previously in (Yeoh et al., 2023), which is originally based on the Fastformer model (Wu et al., 2021). It is imperative to highlight the motivation for proposing TD-SHIFT as an improvement to Non-Local and SHIFT. In contrast to the quadratic complexity of TD-NL, TD-SHIFT has linear computational and memory complexity. This enhancement aims to address the computational challenges associated with TD-NL, ensuring improved performance and scalability in handling temporal focus. This is especially useful for risk prediction with multiple prior screening mammograms to predict cancer risk, as incorporating multiple prior mammograms increases the accuracy of risk prediction but has high memory requirements. For our TD-SHIFT block, only the query is multiplied with the time interval vector, extending time-decay attention to the SHIFT block (Yeoh et al., 2023):

$$SHIFT(Q, K, V, t) = \tilde{k} * V, \tag{9}$$
$$\tilde{Q} = Q * t, \tag{10}$$
$$\tilde{\alpha} = softmax\left(FC_Q(\tilde{Q})\right), \tag{11}$$
$$\tilde{q} = \sum_{j=1}^{\bar{C}} \tilde{\alpha} Q_j, \tag{12}$$
$$p = \tilde{q} * K, \tag{13}$$
$$\tilde{\beta} = softmax(FC_K(p)), \tag{14}$$
$$\tilde{k} = \sum_{j=1}^{\bar{C}} \tilde{\beta} K_j \tag{15}$$

where $FC_Q$ and $FC_K$ are the fully connected layers for attention weights, while $\tilde{q}$ and $\tilde{k}$ are the global query and key. The time interval vector, $t$ only interacts with the query, $Q$ through the Hadamard product. It is crucial to note that the time-decay attention is inherently incorporated in the transformation from query to key. This eliminates the necessity for an explicit Hadamard product of the time interval vector with the key, extending the original Fastformer implementation (Wu et al., 2021). Consequently, the subsequent attention weight calculations through $FC_Q$ and $FC_K$ for both global query and key encapsulate the decaying attention to prior images.

### 3.2.3 Implementation details

We train both attention mechanisms by leveraging weights learned from a previously trained attention model (same architecture) to initialize/jumpstart the learning process. The training process for time-decaying attention starts off at a higher learning rate of 5e$^{-4}$, facilitating rapid adaptation to new inputs. Subsequently, we use a lower learning rate of 5e$^{-5}$ to finetune the model. Experimental results, presented in the Results section, demonstrate the necessity of this approach. These steps are distinct from (Hu et al., 2023), but should achieve the same goal for the different architectures employed. For instance, (Hu et al., 2023) utilizes six different learning rates, throughout the training procedure. However, experimental results showed that our method of using two different learning rates achieved the highest results for our model.

### 3.3 A new Radiomics and Deep learning feature based Multiple Instance Learning (RADMIL) method

In existing literature, explainable methods to integrate radiomic features with deep learning features are lacking, and we recognize the need to advance beyond rudimentary approaches. Conventional methods combine radiomic features with deep learning features by concatenating them (Li et al., 2023; Yeoh et al., 2023). Our previous study and another recent study show that radiomic feature inclusion improves the deep learning network's performance (Li et al., 2023; Yeoh et al., 2023); however, current approaches are rather rudimentary and not interpretable/explainable as to how radiomic features influence deep learning features and vice versa. Rather than treating radiomic features as conventional tabular or clinical data and concatenating them through a FC layer, we propose to consider them as features similar to deep learning features obtained from an encoder-like CNN. This shift in perspective lays the foundation for our proposed approach.

At the heart of this integration is Attention based Multiple Instance Learning (AMIL) (Ilse et al., 2018), which serves as the feature combination method. The usage of AMIL not only enhances the model's predictive capabilities but also improves the interpretability of the model, offering unique insights into each feature's contribution for the risk assessment task. Let $H$ be a bag of $n$ number of features from either deep learning or radiomics, the formulation for AMIL is:

$$z = \sum_{k=1}^{n} a_k h_k, \qquad (16)$$

$$a_k = \frac{exp(FC_2(\tanh(FC_1(h_k))))}{\sum_{j=1}^{n} exp(FC_2(\tanh(FC_1(h_j))))}, \qquad (17)$$

where $FC_1$ and $FC_2$ are trainable parameters in the form of fully connected layers. AMIL pools the features in bag $H$ through weighted averaging using the attention score $a$, calculated to produce a bag-level feature $z$. The compiled feature can then be used to make predictions through an output FC layer.

The attention score $a$, which sums to one, can be used to visualize the contribution of each feature. In the case of deep learning features, it can weigh the importance of the features coming from each CC or MLO view. Through training, the calculated attention scores would facilitate the model in deciding which view contributes more to risk assessment. Importantly, AMIL also allows the model to weigh the importance of deep learning and radiomic features. In our approach, we froze the CNN encoder to preserve the learned weights, ensuring a robust image/feature embedding. However, it is important to note that the results presented here would be comparable if the encoder were replaced by another well-trained encoder.

Inspired by (Wu et al., 2020) that evaluated several ways to combine information from different mammographic views in an exam using FC layers, we explored different combinations of feature combination/integration and compared them to the conventional method of incorporating tabular data through concatenation before the FC layer. The configurations differ in how the features from all 4 views are aggregated to produce the final predictions. The different architectures to combine features from all views are as follows:

1. **Default/Baseline**: Radiomic features are integrated through a FC layer by concatenating with the deep learning features. Predictions are made from each view, and final prediction is obtained by averaging the 4 risk scores.
2. **Config A**: Deep learning and radiomic features are combined through AMIL, creating a bag-level feature for each view. Predictions are made from each view, and the final prediction is obtained by averaging the 4 risk scores.
3. **Config B**: Both deep learning and radiomic features are first combined through an FC layer, forming merged features for each view. These features are then merged using AMIL, and prediction is made using the patient feature.
4. **Config C**: All features from all views are combined directly using AMIL, and predictions are made based on this combined feature.
5. **Config D**: Deep learning features are combined first through AMIL, and then the intermediate bag-level feature is merged with radiomic features through another AMIL. Prediction is made using the final bag-level feature.
6. **Config E**: Deep learning features are combined first through AMIL, and then the intermediate bag-level feature is merged with radiomic features mapped to a lower dimensional space through an FC layer. Prediction is made using the final bag-level feature.

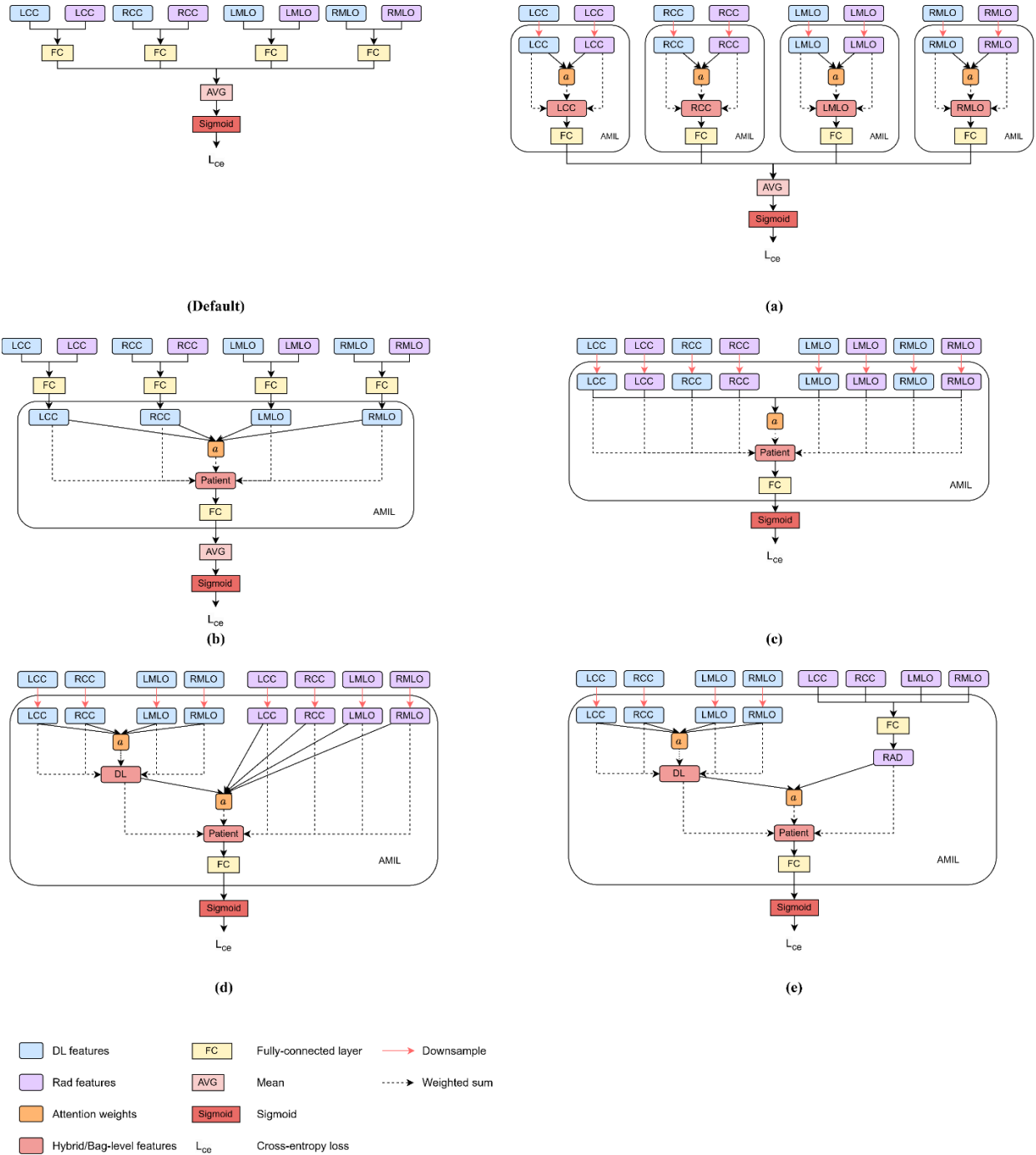

Fig. 6: Different configurations of AMIL implemented through the attention weights, $a$ examined in our network architecture. The Default configuration depicts the conventional/baseline method to combine radiomic (Rad) and deep learning (DL) features that does not incorporate AMIL.

### 3.3.1 Integration of lateral attention for improved risk assessment

As cancer usually develops in/only affects one side of the breast, having the ability to identify the cancer laterality should be beneficial for early detection. We hereby extend the initial AMIL configuration to account for cancer laterality, thus providing additional interpretability to the model.

Drawing insights from (Huang et al., 2023), we introduce an additional attention score called lateral attention $l$, which is trained in a supervised manner. In contrast to the softmax attention $a$ from the previous section, the lateral attention scores serve a different purpose. Instead of uniformly aggregating features from both breasts, the lateral attention is specifically designed to steer the attention mechanism to focus more on the affected breast. Incorporating lateral attention results in a modification of the original AMIL equation (i.e., equation 16), as follows:

$$z = \sum_{k=1}^{n} \bar{a}_k h_k, \tag{18}$$

$$\bar{a} = \frac{a_k l_k}{\sum_i^n a_i l_i}, \tag{19}$$

$$l_k = \sigma(FC_{l2}(\tanh(FC_{l1}(h_k)))), \tag{20}$$

where $FC_{l1}$ and $FC_{l2}$ are trainable parameters for lateral attention and $\sigma$ represents either the sigmoid or softmax function. With this formulation, the full lateral attention $l_k$ that ranges from 0-1 can be given to a relevant (e.g., cancer affected) image, whereas the classifier will learn to allocate an attention of near 0 for an image that is irrelevant (e.g., a normal/unaffected breast).

We utilized the cancer (left/right) laterality information of the histopathological results from the datasets to construct both soft and hard training labels. Our experiments (see Results section) showed that Config E was the best architecture to aggregate deep learning and radiomic features; therefore, we extended Config E to incorporate lateral attention as displayed in Fig. 7.

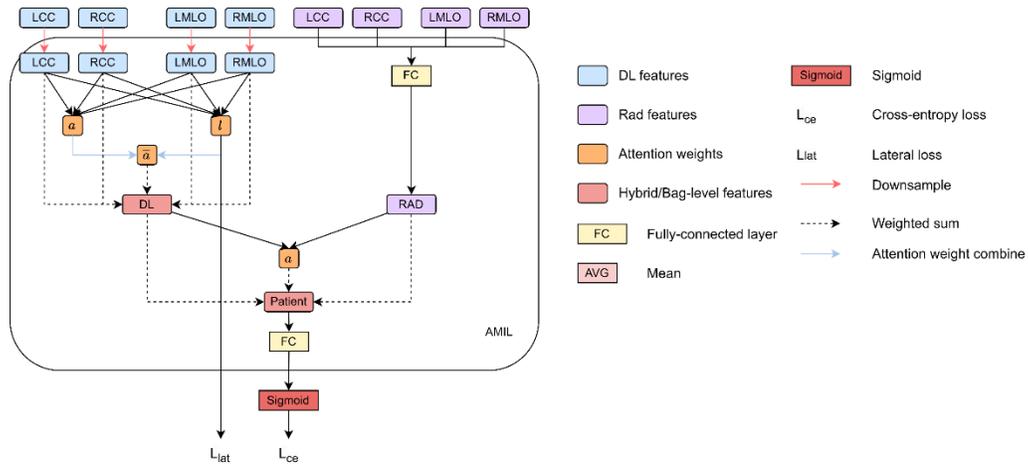

Fig. 7: Integration of lateral attention with Config E of Fig. 6.

In our experiments, we found that simultaneously training both softmax attention $a$ in AMIL and lateral attention $l$ produced a performance drop, which could be due to task interference as the learning of one task inhibited that of the other. To circumvent this issue, we first trained our AMIL architecture without lateral attention. Following the successful training of AMIL, we utilized the learned weights as initialization; by keeping the AMIL weights frozen, we exclusively trained the lateral attention component. This approach aimed to leverage the knowledge acquired by AMIL and prevented interference between the two components during training.

### 3.4 ReST$^{CL}$: Reinforced Self-Training with Continual Learning

Recent advancements in machine learning have demonstrated the potential of self-training frameworks, such as Reinforced Self-Training (ReST) (Gulcehre et al., 2023) for improved classification performance. The ReST method and its variant with expectation-maximization (ReST$^{EM}$) (Singh et al., 2023) improve model performance using self-generated data. ReST leverages a combination of synthetic data and human-labeled data and refines the model through iterative finetuning.

Here, we propose a new ReST approach called ReST$^{CL}$ that uses a continual learning (CL) (Wang et al., 2023) framework to further improve an existing risk model on a secondary dataset, which may come from a different population. Unlike (Singh et al., 2023; Xie et al., 2020), we don't filter out samples based on a confidence threshold or a reward function. ReST$^{CL}$ allows the model to incrementally adapt to new data from other populations without catastrophic forgetting or model performance degradation on the original population. The learning procedure can be repeated as new datasets become available, enabling the model to continually learn whilst avoiding model collapse (Shumailov et al., 2023).

In the original ReST method (Gulcehre et al., 2023), the authors start off by training an initial model $\pi_\theta(y|x)$ to map inputs $x$ to outputs $y$ on a given dataset of sequence pairs $D_P$ using the negative log likelihood loss, $L(x, y; \theta)$. A "grow" step creates a new dataset $D_g$, which augments the initial training dataset with samples/synthetic data from the model:

$$D_g = \left\{(x^i, y^i)|_{i=1}^{N_g} \text{ such that } x^i \sim D_P, y^i \sim \pi_\theta(y|x^i)\right\} \cup D_P \tag{21}$$

Subsequently, an "improve" step uses $D_g$ to finetune the policy $\pi_\theta$. The authors first define a filtering function that includes only samples with rewards, $R(x, y)$ higher than a certain threshold $\tau$:

$$F(x, y; \tau) = \mathbb{1}_{R(x,y) > \tau} \tag{22}$$

Next, the authors finetune the current best policy typically trained with an offline reinforcement learning loss or the supervised learning loss $L(x, y; \theta)$ on the filtered data. To sum up, they use the following reward weighted loss $J$:

$$J(\theta) = \mathbb{E}_{(x,y) \sim D_g}[F(x, y; \tau) L(x, y; \theta)] \tag{23}$$

When iterating over "improve" steps, the authors increase the filtering thresholds: $\tau_1 < \cdots < \tau_{N-1} < \tau_N$. The expression for the gradient takes the following form:

$$\nabla J(\theta) = -\mathbb{E}_{x \sim D}\left[\lambda \mathbb{E}_{y \sim \pi_{\theta'}(y|x)}[F(x, y; \tau) \nabla \log \pi_\theta(y|x)] \\ + (1 - \lambda) \mathbb{E}_{y \sim p(y|x)}[F(x, y; \tau) \nabla \log \pi_\theta(y|x)]\right] \tag{24}$$

with $\theta'$ being the parameters of the model from the last "grow" step, $\lambda$ the proportion of data sampled from this model in $D_g$ and a single step of growth. The authors state that the second term in (24) is a form of offline policy gradients that prevents $\pi_\theta(y|x)$ to move too far from $p(y|x)$, which could lead to model collapse (Shumailov et al., 2023).

However, there are several issues with the ReST model that we hope to address here: First, the reward function $R(x, y)$ is not explicitly stated; the authors only state that they use reference-free reward models. Second, the authors' claim that their model does not suffer from model collapse might not be entirely correct as the filtering function in equations (22) and (23) applies to the grown dataset $D_g$, which is a combination of the original/primary dataset $D_P$ and the new/synthetic samples. In other words, the primary dataset $D_P$ gets slashed/reduced in subsequent training/iteration steps, which could lead to the model drifting further and further away from the initial population distribution/training dataset, leading to model collapse. Third, the authors picked the filtering thresholds $\tau_i$ from a sequence of increasing values: [0.0, 0.7, 0.8, 0.9, 0.95, 0.99]. These seem to be rather arbitrarily decided, and could there be a better way of filtering dataset samples that could improve classification performance?

To address these issues, we present our new ReST$^{CL}$ method. In our application, the secondary dataset is the CSAW dataset that originates from a Swedish population. For each exam date, there is an assessment by a first radiologist, a second radiologist, and a consensus decision. In this case, the reward model and filtering function is clear, and is based on the radiologists' consensus decision. For the American EMBED dataset, a semi-automated supervised machine learning pipeline and a hierarchical hybrid natural language processing (NLP) system are used to extract pathologic diagnoses from free-text pathology reports. In a way, the filtering function of the CSAW dataset based on radiologists' consensus might be somewhat analogous to Reinforcement learning from human feedback (RLHF) used to improve the quality of large language model's (LLM) outputs by aligning them with human preferences.

To address the second issue of ReST, we propose a new approach for ReST$^{CL}$ that enables continual learning of the model without slashing the original dataset $D_P$, thus avoiding model collapse. To address the third issue, we filter new samples from the secondary dataset not by applying arbitrary thresholds, but by stratifying samples based on bilateral asymmetry characteristics related to cancer risk in the finetuning procedure/process. Given that the secondary dataset comes from a different population, we implemented a mixed-labeling strategy to assign either hard true labels or soft pseudo-label to the samples. The choice between hard and soft labels is guided by the lateral attention scores assigned to each view by the model trained on the primary EMBED dataset. We determine the model's confidence in a new sample by calculating the difference in attention scores between the left and right breasts. A case with a high difference in attention scores signifies strong confidence by the model; likewise, for a control with a low difference in attention scores. Since cancer typically develops in only one breast, cases typically demonstrate higher bilateral asymmetry between left and right breasts compared to controls. Thus, the difference in lateral attention score is formulated as:

$$\begin{aligned}\Delta A(x) = \ &|A(x_L) - A(x_R)| \\ = \ &|(A(x_{LCC}) + A(x_{LMLO})) - (A(x_{RCC}) + A(x_{RMLO}))|\end{aligned} \quad (25)$$

Thus, for the secondary CSAW dataset, a hard label is assigned if the lateral difference exceeds the 99$^{th}$ quantile $Q_{99}^{case}$ for cases. For controls, a hard label is assigned if the lateral difference is below the 1st quantile $Q_1^{control}$. As the model has high confidence in these samples, it is more likely that the model's prediction aligns with the true label. Training on these confident samples with hard labels reinforces correct decision boundaries, enhancing the overall accuracy of the model. In this way, the label assignment function $Lab(x)$ is defined as:

$$Lab(x) = \begin{cases} \text{Hard label, if } \Delta A(x) \subseteq Q_{99}^{case} \cup Q_{1}^{control} \\ \text{Soft pseudo label, if } \Delta A(x) \not\subseteq Q_{99}^{case} \cup Q_{1}^{control} \end{cases} \quad (26)$$

This improvement is also achieved by retaining $D_P$, which ensures that the model does not lose its ability to generalize to the original population while gaining insights from new data sources. We also first finetune on the secondary dataset, before going back to finetuning on our original dataset of the target population in each epoch during training. This alternating process ensures that the model learns from new data without experiencing catastrophic forgetting, preserving performance on the target population. Algorithm 1 outlines the full ReST$^{CL}$ algorithm.

---

**Algorithm 1: ReST$^{CL}$ algorithm.** ReST$^{CL}$ promotes continual learning using laterality based label assignments. The initial classifier is trained on the primary dataset. Then, ReST$^{CL}$ iteratively applies finetuning on the secondary dataset and finetuning on the primary dataset in an alternating process to update the classifier.

---

**Input**:
- Primary Dataset $D_P$: Original dataset from the target population, with labeled samples.
- Secondary Dataset $D_S$: New dataset from a different population, with labeled samples.
- Batch Sizes $B_P$, $B_S$: Batch sizes for primary and secondary datasets.
- $E$: Number of epochs.
- $L_S$: Loss on secondary dataset.
- $L_P$: Loss on primary dataset.

**Initialize**:
- Set thresholds for label assignment: case threshold $Q_{99}^{case}$ and control threshold $Q_{1}^{control}$.
- Train $\pi_\theta$ on $D_P$ using loss $L_P$

**for** $e = 1$ to $E$ **do**:
    // Fine-tune on secondary dataset
    **for** mini batch $B_S$ from $D_S$ **do**:
        **for** each sample in $B_S$ **do**:
            Compute lateral difference: $\Delta A(x) = |A(x_L) - A(x_R)|$
            **if** $\Delta A(x) \subseteq Q_{99}^{case} \cup Q_{1}^{control}$:
                Assign hard label
            **else**:
                Assign soft pseudo-label
            end
        end
        Compute loss $L_S$ & Update model parameters
    **end**

    // Fine-tune on primary dataset
    **for** mini batch $B_P$ from $D_P$ **do**:
        Compute loss $L_P$ & Update model parameters
    **end**
**end**

**Output**: Classifier $\pi_\theta$

## 3.5 Cancer forecast network with time-interval embeddings

Developing risk forecast models that empower radiologists to recommend personalized screening programs would optimize resource allocation, reduce radiation accumulation in a woman's body, and help implement preventative measures for early cancer detection. By leveraging deep learning techniques and incorporating various risk factors, these models can provide a dynamic assessment of an individual's likelihood of developing cancer throughout a period. This approach helps to stratify high-risk patients from low-risk individuals and facilitate early intervention.

The additive hazard layer introduced in (Yala et al., 2021) aims to predict cancer risk by utilizing mammographic features and traditional risk factors. The hazard layer predicts a patient's risk for each year over the next 5 years by first predicting a baseline risk using a small linear/FC layer, $B(x)$. This baseline risk is the cancer risk of the individual in the current year, which is the standard risk score of other risk models in the literature. The additive hazard layer subsequently calculates a marginal hazard for each year separately using individual networks, each implemented as a linear layer followed by a ReLU activation function denoted as $H_i(x)$. The overall risk at year $\bar{T}$ is obtained by summing the baseline risk and the marginal hazards up to that particular year (i.e., year 1 to 5 for a 5-year forecasting model) (Yala et al., 2021), as follows:

$$P(T_{cancer} = \bar{T} \mid x) = B(x) + \sum_i^{\bar{T}} H_i(x) \tag{27}$$

The additive property of the hazard layer ensures that a patient's risk in two years' time is always higher than their risk in the first year while offering a clear visual representation of risk trajectories, aiding in timely clinical interventions. Cancer patients will have risk scores that escalate significantly over time as compared to control patients who maintain a steady, low-risk profile.

By building upon the additive hazard layer, we propose a new method incorporating an embedding layer to use time intervals between patients' previous and current mammographic screenings for improved predictive accuracy. Adding this time-interval embedding helps the model adjust the risk scores more precisely by adding a temporal dimension for better risk prediction or classification in the feature space. The time-interval input embedding ranges from 0 to 10, representing six-month intervals up to 5 years (i.e., 0 represents cancer occurrence in 0 years; 1 represents cancer occurring in 6 months' time; 2 represents cancer occurring in 1 years' time, etc.). A similar idea was first presented in (Hu et al., 2023) for a glaucoma forecast network. However, in that paper (Hu et al., 2023), the authors input the time interval as a token into a transformer network. As glaucoma screening intervals are not fixed, unlike breast cancer screening paradigms, which typically occur at regular intervals (i.e., 6 months, 1 year, or 2 years), the authors included the time interval as an additional token to the input of an image-based encoder. However, many issues with tokenization, including multi-modality tokenization (Spathis and Kawsar, 2024) have led researchers to try to eliminate this process altogether (Yu et al., 2023). Our method of incorporating a time-interval embedding is similar to providing the risk model labels or context in a conditional generative adversarial network (CGAN) (Mirza and Osindero, 2014), to help the model to forecast individualized screening intervals better.

To the best of our knowledge, the inclusion of the time-interval embeddings has not been examined yet in the literature (Arefan et al., 2019; Dadsetan et al., 2022; Lee et al., 2023a; Yala et al., 2021), which could be crucial information to guide the model towards better prediction in the feature space. Yala et al. (Yala et al., 2022) proposed the Tempo-Mirai reinforcement learning based method to recommend personalized screening intervals. However, training neural network reinforcement learning methods is challenging (Ding and Dong, 2020b) due to sample efficiency issues, training stability, etc. (Ding and Dong, 2020a), which might be unnecessary for a fixed time-interval screening paradigm. Furthermore, our approach is simple, enabling radiologists to recommend screening recommendations based on the operating points along the ROC curves for each time interval (6 months, 1 year, 1.5 years, etc.). The formulation for the new additive layer incorporating time-interval embeddings is thus:

$$P(T_{cancer} = \bar{T} \mid x) = B(x) + \sum_i^{\bar{T}} H_i(x + e(t)) \quad (28)$$

where $e(t)$ represents the embedding features of time interval, $t$ between screenings and $\bar{T}$ is the time of cancer diagnosis. We incorporate this new additive layer into our lateral-attentive AMIL based method as depicted in Fig. 8. The $x$ input in the additive layer represents the aggregated deep learning features from all four views with the radiomic features. The additive layer can thus be fully expressed as follows:

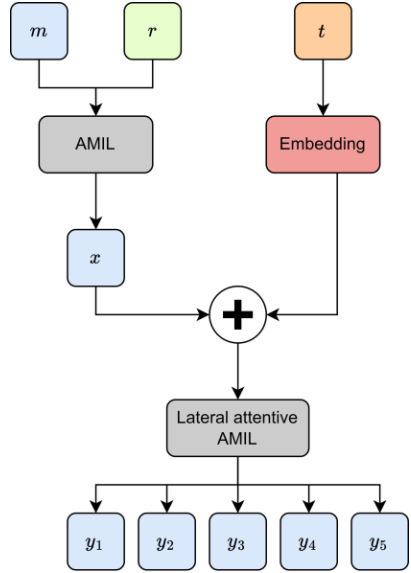

Fig. 8: Additive hazard layer with time-interval screening embeddings. The deep learning features $m$ are combined with radiomic features $r$ by our lateral-attentive AMIL based method. Then, the resulting feature embedding $x$ is added with the screening time-interval $t$ embedding, to form an additive hazard for future cancer risk prediction.

$$P(T_{cancer} = \bar{T} \mid m, r) = B(\text{AMIL}(m, r)) + \sum_i^{\bar{T}} H_i(\text{AMIL}(m, r) + e(t)) \quad (29)$$

where $m$ represents deep learning features obtained from the CNN encoder and $r$ represents the extracted radiomic features. AMIL represents our lateral-attentive AMIL based method presented in the previous section.

Our new method is not just constrained to predicting future cancer risk of screening mammograms; it can be used in any forecasting network that implements screening at regular time intervals. Namely, the time-interval embedding can be incorporated for any risk model that involves screening at regular intervals – in our study, 6-month intervals were applicable, but the method can also be used for risk prediction of annual lung cancer screening in computed tomography (CT) scans, for example.

### 3.6 Experimental setup and classification methodology

#### 3.6.1 Preprocessing

We maintained the same preprocessing steps as our previous study (Yeoh et al., 2023). First, we used Otsu's thresholding to segment the images, removing background regions and irrelevant labels. Each image was resized to $256 \times 256$ and padded to maintain the aspect ratio. We added prior screening images by stacking the processed mammograms in chronological order, with each screening as a separate time point in a video-like sequence.

We also normalized each image to have zero mean and unit variance by calculating the mean and standard deviation within each cross-validation fold's training subset. To introduce regularization and prevent overfitting, we applied random augmentations during training, including horizontal and vertical flips with a 0.5 probability.

#### 3.6.2 Implementation details

We used the Adam (Kingma and Ba, 2015) optimizer to train our model with the default parameters of $\beta_1 = 0.9$, $\beta_2 = 0.999$ and $\epsilon = 1e-8$. The learning rate was set to $1e^{-4}$ and batch size of 12.

During continual learning with label assignment, the vanilla SGD (Ruder, 2016) optimizer was used instead for finetuning the model, with a significantly reduced learning rate of $1e^{-7}$. This switch from the Adam optimizer to vanilla SGD was chosen to facilitate more stable, incremental updates during the continual learning phase, which was critical for preserving previously learned knowledge while integrating new information without overfitting.

Our entire code, including CNN architecture and analysis scripts, is implemented in Python (v3.8.5) using Scikit-learn (v1.0.2), Matplotlib (v3.5.1) and PyTorch (v1.10.1). We implemented our TRINet model on 2 RTX 2080 Ti GPUs, each with 12 GB of memory.

#### 3.6.3 Performance metrics

We calculated the 1-year to 5-year Area Under the receiver operating Curve (AUC) using the same approach as in (Yala et al., 2021). For example, to compute the 3-year AUC, we considered as positive cases that had a cancer diagnosis within 3 years of a previous year's mammogram. We computed the 1-year, 2-year to 5-year AUC in the same way as per the ground truth labels in the dataset.

All results are tabulated in the Results section. We also utilized bootstrapping methods to obtain the 95% confidence intervals (CI) for all the AUC values. ROC curves are also plotted for the 1-year to 5-year prediction results to measure the true positive rates versus the false positive rates for each prediction category. We also performed ablation studies of each novelty introduced in the Methods section to demonstrate the improvement brought about by each new addition to the model.

## 4. Results
### 4.1 Time-Decay attention

Table 1 presents the results of different attention mechanisms integrated into our CNN encoder and their time-decay counterparts. As shown in our previous work (Yeoh et al., 2023), the inclusion of attention mechanisms into the deep learning model can greatly improve risk prediction performance. The results in Table 1 demonstrate the benefits of including information in the time dimension in both Non-Local and SHIFT attention mechanisms. A time-sensitive self-attention mechanism in (Hu et al., 2023) called GLIM is also compared using Non-Local as the baseline method.

Specifically, we observe that the introduction of time-decay attention between screenings contributes to improved model performance across almost all metrics. We observe the highest performance improvement in 1- and 2-year AUCs, possibly attributed to more information present in recent screenings as opposed to previous screenings for recent years' cancer prediction. Both time-decay versions of Non-Local and SHIFT attention mechanisms outperform GLIM. We attribute this to our approach of performing the Hadamard product directly on the query and key vectors with the time vector. This direct scaling of the query and key vectors by the time vector enhances the temporal sensitivity of the attention mechanism more effectively than applying the time matrix to the query-key product.

Table 1: Summary of AUC results with 95% confidence intervals (CIs) for GLIM (Hu et al., 2023), Non-Local networks, SHIFT, and their respective time-decay variants. An attention block version of GLIM is used as a comparison. The "+ T" here denotes the time-decay variants.

| Attention Model | 1-year AUC | 2-year AUC | 3-year AUC | 4-year AUC | 5-year AUC |
|---|---|---|---|---|---|
| Baseline | 0.789 (0.746-0.835) | 0.756 (0.713-0.799) | 0.747 (0.704-0.787) | 0.737 (0.696-0.780) | 0.739 (0.697-0.777) |
| NL | 0.805 (0.763-0.847) | 0.775 (0.730-0.817) | 0.765 (0.724-0.808) | 0.752 (0.713-0.795) | 0.753 (0.713-0.794) |
| NL + T | 0.821 (0.782-0.864) | 0.784 (0.743-0.831) | 0.765 (0.727-0.805) | 0.757 (0.718-0.797) | 0.756 (0.717-0.796) |
| GLIM (Hu et al., 2023) | 0.813 (0.774-0.855) | 0.783 (0.742-0.827) | **0.772** (0.733-0.811) | 0.757 (0.718-0.798) | 0.757 (0.719-0.798) |
| SHIFT | 0.81 (0.769-0.852) | 0.777 (0.737-0.819) | 0.771 (0.735-0.816) | 0.755 (0.716-0.797) | 0.754 (0.717-0.796) |
| SHIFT + T | **0.825** (0.783-0.867) | **0.784** (0.742-0.829) | 0.771 (0.729-0.816) | **0.764** (0.725-0.805) | **0.76** (0.721-0.8) |

The results in Table 2 show that models trained with the two-step learning rate strategy consistently outperformed those trained with a single learning rate of $1e^{-5}$. For instance, Non-Local trained with the two-step learning rate achieves a 1-year AUC of 0.821 compared to 0.813 with a single learning rate. Similarly, SHIFT achieves a 1-year AUC of 0.825 with the two-step strategy versus 0.816 with the single learning rate. These AUC improvements reinforce the necessity of these steps to handle the domain shift for the new time vector inputs, which may imitate the training approach from (Hu et al., 2023) that cycles through 6 different learning rates to handle their time matrix.

To effectively apply time-decay attention, the model requires pretrained weights from a previously trained attention model as an initial starting point. For example, we first train a model with Non-Local attention, then we use this trained model to finetune a time-decay attention model Table 3 highlights the impact of initializing time-decay attention mechanisms with weights from previously trained attention models. Moreover, the performance of the time-decay models without pre-trained weights is even worse than the baseline Non-Local and SHIFT models. For instance, the time-decay Non-Local model without pre-trained weights achieves a 1-year AUC of 0.746, which is significantly lower than 0.805 achieved by the baseline Non-Local model. This degradation in performance can be attributed to the time vector obfuscating the training process, hindering the model's ability to converge without a good initial set of attention weights.

Table 2: AUC results with 95% CIs of training different time-decay attention mechanisms using different finetuning learning rates.

| Attention Model | Learning Rates | 1-year AUC | 2-year AUC | 3-year AUC | 4-year AUC | 5-year AUC |
|---|---|---|---|---|---|---|
| NL+T | [1e-5] | 0.813 (0.774-0.855) | 0.782 (0.744-0.825) | **0.772** (0.733-0.81) | 0.756 (0.717-0.795) | 0.757 (0.718-0.795) |
| | [5e-4, 5e-5] | 0.821 (0.782-0.864) | 0.784 (0.743-0.831) | 0.765 (0.727-0.805) | 0.757 (0.718-0.797) | 0.756 (0.717-0.796) |
| SHIFT + T | [1e-5] | 0.816 (0.776-0.86) | 0.778 (0.736-0.82) | 0.759 (0.713-0.799) | 0.743 (0.699-0.785) | 0.741 (0.701-0.786) |
| | [5e-4, 5e-5] | **0.825** (0.783-0.867) | **0.784** (0.742-0.829) | 0.771 (0.729-0.816) | **0.764** (0.725-0.805) | **0.76** (0.721-0.8) |

Table 3: AUC results with 95% CIs of training different time-decay attention mechanisms using initialized weights from previously trained attention models.

| Attention Model | Initialized Weights | 1-year AUC | 2-year AUC | 3-year AUC | 4-year AUC | 5-year AUC |
|---|---|---|---|---|---|---|
| NL+T | - | 0.746 (0.7-0.795) | 0.729 (0.686-0.773) | 0.722 (0.681-0.766) | 0.706 (0.662-0.749) | 0.708 (0.667-0.75) |
| | ✓ | 0.821 (0.782-0.864) | 0.784 (0.743-0.831) | 0.765 (0.727-0.805) | 0.757 (0.718-0.797) | 0.756 (0.717-0.796) |
| SHIFT + T | - | 0.793 (0.75-0.836) | 0.755 (0.716-0.8) | 0.739 (0.696-0.783) | 0.730 (0.691-0.772) | 0.728 (0.69-0.77) |
| | ✓ | **0.825** (0.783-0.867) | **0.784** (0.742-0.829) | **0.771** (0.729-0.816) | **0.764** (0.725-0.805) | **0.76** (0.721-0.8) |

To find the optimal values for parameters *A*, *B* and threshold *T* in the time-decay attention (equation 8), we performed a hyperparameter sweep of possible values, similar to (Hu et al., 2023). The results of the different hyperparameter values are tabulated in Table 4, 5 and 6. From these tables, we observe that the best performance was achieved with $A = 2.0$ and $B = 0.1$, whereby the model maintained high AUC scores across all performance metrics, reflecting an effective balance in how past data influences predictions at these parameter values. Lower values of *A* and *B* led to insufficient attention decay, under-prioritizing recent data, while higher values caused excessive decay, leading to the underutilization of valuable historical information.

Table 4: AUC results with 95% CIs for different values of *A* with a fixed value of $B = 0.1$ and threshold, $T = 60$ for SHIFT+T attention.

| *A* | 1-year AUC | 2-year AUC | 3-year AUC | 4-year AUC | 5-year AUC |
|---|---|---|---|---|---|
| 1.6 | 0.813 (0.773-0.858) | 0.780 (0.738-0.825) | 0.769 (0.731-0.811) | 0.763 (0.725-0.8) | 0.763 (0.726-0.803) |
| 1.8 | 0.818 (0.776-0.864) | 0.781 (0.737-0.827) | 0.762 (0.717-0.804) | 0.747 (0.704-0.788) | 0.743 (0.702-0.785) |
| 2 | **0.825** (0.783-0.867) | 0.784 (0.742-0.829) | 0.771 (0.729-0.816) | **0.764** (0.725-0.805) | **0.76** (0.721-0.8) |
| 2.2 | 0.817 (0.777-0.857) | **0.785** (0.745-0.826) | **0.773** (0.736-0.81) | 0.762 (0.726-0.803) | 0.760 (0.725-0.8) |
| 2.4 | 0.816 (0.772-0.859) | 0.782 (0.741-0.825) | 0.765 (0.723-0.811) | 0.748 (0.71-0.789) | 0.746 (0.708-0.791) |

Table 5: AUC results with 95% CIs for different values of $B$ with a fixed value of $A = 2.0$ and threshold, $T = 60$ for SHIFT+T attention.

| $B$ | 1-year AUC | 2-year AUC | 3-year AUC | 4-year AUC | 5-year AUC |
|---|---|---|---|---|---|
| 0.01 | 0.812 (0.772-0.857) | 0.775 (0.736-0.822) | 0.758 (0.718-0.802) | 0.744 (0.702-0.784) | 0.743 (0.704-0.786) |
| 0.05 | 0.803 (0.764-0.843) | 0.778 (0.734-0.821) | 0.770 (0.731-0.814) | 0.760 (0.723-0.798) | 0.757 (0.724-0.796) |
| 0.1 | **0.825** (0.783-0.867) | **0.784** (0.742-0.829) | **0.771** (0.729-0.816) | **0.764** (0.725-0.805) | **0.76** (0.721-0.8) |
| 0.5 | 0.811 (0.773-0.851) | 0.776 (0.737-0.82) | 0.754 (0.711-0.798) | 0.741 (0.701-0.788) | 0.739 (0.699-0.781) |
| 1 | 0.791 (0.748-0.838) | 0.758 (0.714-0.802) | 0.745 (0.699-0.787) | 0.732 (0.692-0.775) | 0.730 (0.687-0.771) |

Table 6 tabulates the effects of different threshold values, $T$ used in the time-decay attention. Time-decay attention performs best at $T = 60$ (i.e., 5 years); performance deterioration occurs for other values of $T$.

Table 6: AUC results with 95% CIs for different values of threshold, $T$ with a fixed value of $A = 2.0$ and $B = 0.1$ for SHIFT+T attention.

| $T$ | 1-year AUC | 2-year AUC | 3-year AUC | 4-year AUC | 5-year AUC |
|---|---|---|---|---|---|
| 48 | 0.819 (0.781-0.857) | **0.786** (0.748-0.832) | 0.768 (0.741-0.82) | 0.754 (0.731-0.805) | 0.750 (0.728-0.803) |
| 54 | 0.823 (0.781-0.863) | 0.783 (0.749-0.832) | 0.764 (0.738-0.816) | 0.748 (0.725-0.805) | 0.744 (0.729-0.803) |
| 60 | **0.825** (0.783-0.867) | 0.784 (0.742-0.829) | **0.771** (0.729-0.816) | **0.764** (0.725-0.805) | **0.76** (0.721-0.8) |
| 66 | 0.818 (0.779-0.866) | 0.777 (0.733-0.822) | 0.756 (0.715-0.799) | 0.743 (0.701-0.787) | 0.740 (0.697-0.782) |
| 72 | 0.808 (0.771-0.854) | 0.779 (0.739-0.819) | 0.766 (0.729-0.806) | 0.752 (0.709-0.792) | 0.752 (0.714-0.792) |

### 4.2 Radiomics integration with RADMIL

A results summary of integrating radiomic features using different RADMIL configurations is tabulated in Table 7. From the results of Config C with and without radiomics feature inclusion in Table 7, we make two important observations: First, we observe that radiomic feature inclusion using our new attention based RADMIL method improves model performance. Similar performance improvements were observed for other configurations. Second, we observe the importance of using attention to aggregate features from all four images/views as each image contributes uniquely to risk prediction.

The lower results of Configs A and B show that combining radiomic features early with their corresponding deep learning features does not yield significant improvements in the baseline result. We conclude that the best RADMIL methods first aggregate the features from deep learning and radiomics individually. The aggregated result should be combined subsequently using another AMIL mechanism as implemented in Configs E and D. The highest results were obtained with Config E, which combined radiomic features with an FC layer and deep learning features with AMIL, followed by combining both results with another AMIL mechanism.

Table 7: AUC results with 95% CIs for the integration of radiomic features with deep learning features using different configurations of RADMIL.

| RADMIL Configuration | 1-year AUC | 2-year AUC | 3-year AUC | 4-year AUC | 5-year AUC |
|---|---|---|---|---|---|
| A | 0.832 (0.784-0.881) | 0.788 (0.738-0.84) | 0.779 (0.733-0.833) | 0.775 (0.731-0.822) | 0.772 (0.726-0.822) |
| B | 0.834 (0.793-0.88) | 0.795 (0.748-0.844) | 0.790 (0.746-0.834) | 0.787 (0.748-0.829) | 0.786 (0.746-0.827) |
| C | 0.847 (0.802-0.895) | 0.804 (0.757-0.853) | 0.791 (0.743-0.84) | 0.787 (0.745-0.831) | 0.784 (0.74-0.83) |
| C (no radiomics) | 0.845 (0.798-0.893) | 0.800 (0.751-0.849) | 0.788 (0.743-0.838) | 0.786 (0.742-0.831) | 0.783 (0.74-0.83) |
| D | 0.850 (0.8-0.901) | 0.804 (0.752-0.856) | 0.793 (0.746-0.843) | 0.790 (0.745-0.833) | 0.787 (0.744-0.836) |
| E | **0.852** (0.808-0.901) | **0.807** (0.756-0.859) | **0.795** (0.751-0.842) | **0.791** (0.747-0.837) | **0.788** (0.746-0.833) |

Table 8 tabulates various methods of combining additional features into the deep learning model using (i) a conventional FC layer, (ii) RADMIL and (iii) RADMIL combined with lateral attention. The results demonstrate that our RADMIL method, both with and without the addition of lateral attention, consistently outperforms other methods. Despite the relatively small gains from lateral attention, its inclusion consistently outperforms RADMIL alone on 4 out of 5 metrics suggesting underlying benefits in its inclusion to enhance the model's feature representation capabilities. These results illustrate that while simpler methods like concatenating additional features through a conventional FC layer can offer decent performance, more sophisticated approaches like AMIL significantly enhance prediction accuracy, validating the importance of advanced integration techniques in deep learning models.

Table 8: AUC results with 95% CIs for the integration of radiomic features using different combination methods. Configuration E of RADMIL is used here as it produces the highest results in Table 7.

| Feature Combination Method | 1-year AUC | 2-year AUC | 3-year AUC | 4-year AUC | 5-year AUC |
|---|---|---|---|---|---|
| FC | 0.833 (0.789-0.877) | 0.794 (0.745-0.846) | 0.791 (0.745-0.839) | 0.788 (0.748-0.831) | 0.786 (0.744-0.831) |
| RADMIL | **0.852** (0.808-0.901) | 0.807 (0.756-0.859) | 0.795 (0.751-0.842) | 0.791 (0.747-0.837) | 0.788 (0.746-0.833) |
| RADMIL + Lat | 0.851 (0.806-0.904) | **0.811** (0.764-0.862) | **0.796** (0.751-0.839) | **0.793** (0.751-0.84) | **0.789** (0.743-0.831) |

### 4.3 Continual learning with ReST$^{CL}$

Table 9 tabulates the results of utilizing CSAW as a secondary dataset for continual learning. The baseline result in the table reflects the performance of the model using a modified ReST and ReST$^{EM}$ approach of finetuning on both CSAW and EMBED datasets. The reward function in ReST is replace with confidence threshold similar in the paper (Gulcehre et al., 2023). The ReST$^{EM}$ finetuning result is considerably lower than the two ReST$^{CL}$ methods indicating that the inclusion of an additional dataset using our new approach can benefit an existing model even though the new dataset's population/cohort is considerably different from the original dataset's population.

The results also show that our method of label assignment based on lateral attention improves the finetuning procedure/process. The improvements suggest that reassigning labels with a focus on laterality helps our model better capture bilateral asymmetry in the data, which is important for risk assessment, confirming the results in Table 8. The second iteration of label assignment maintains the improvements observed after the first iteration, confirming the stability of the approach. Only a slight improvement is observed in the 2-year AUC result after the second iteration of ReST$^{CL}$, which indicates that the model has reached a plateau in performance after this iteration.

Table 9: AUC results with 95% CIs for continual learning using the CSAW dataset as a secondary dataset for the original ReST method, ReST$^{CL}$ and ReST$^{EM}$.

| Learning Method | 1-year AUC | 2-year AUC | 3-year AUC | 4-year AUC | 5-year AUC |
|---|---|---|---|---|---|
| ReST (Threshold: 0.7) | 0.846 | 0.808 | 0.796 | 0.793 | 0.790 |
| ReST (Threshold: 0.35) | 0.853 | 0.812 | 0.797 | 0.792 | 0.789 |
| ReST$^{EM}$ (finetuning on EMBED and CSAW) | 0.8537 (0.811-0.903) | 0.8132 (0.766-0.865) | 0.8007 (0.758-0.847) | 0.7965 (0.759-0.84) | 0.7929 (0.751-0.834) |
| ReST$^{CL}$: (1$^{st}$ Iteration) | **0.8549** (0.815-0.904) | 0.8139 (0.769-0.863) | **0.8014** (0.759-0.851) | **0.7971** (0.754-0.841) | **0.7934** (0.752-0.838) |
| ReST$^{CL}$: (2$^{nd}$ Iteration) | 0.8549 (0.808-0.901) | **0.8140** (0.768-0.862) | 0.8014 (0.757-0.848) | 0.7970 (0.756-0.84) | 0.7934 (0.752-0.837) |

### 4.4 Cancer forecast prediction with time-interval embeddings

Using the additive hazard layer from (Yala et al., 2021), we can forecast a patient's probability of developing breast cancer in 1 to 5 years' time, as an ever-increasing risk progression. This is notably different from the previous models that generate a single risk score, which is then used to forecast/compute all five 1- to 5-year AUC scores (Yeoh et al., 2023). Table 10 tabulates the results of augmenting the output layer of our model with a time-interval embedding and an additive hazard layer.

In Table 10, the introduction of a time-interval embedding into the additive hazard layer improves most of the metrics namely, the 1 to 3-year AUC results. In the EMBED dataset, there is a much higher number of patients with only 1 to 3 consecutive screenings compared to 4 and 5 total screenings (please refer to Fig. 3); thus, the 1 to 3-year AUC results might be a more accurate reflection of the results in Table 10.

Table 10: AUC results with 95% CIs of risk prediction using an additive hazard layer and additive hazard layer combined with time embeddings.

| Forecast Prediction Method | 1-year AUC | 2-year AUC | 3-year AUC | 4-year AUC | 5-year AUC |
| --- | --- | --- | --- | --- | --- |
| Additive Hazard (Yala et al., 2021) | 0.857 (0.821-0.895) | 0.814 (0.776-0.853) | 0.796 (0.759-0.838) | **0.783** (0.744-0.821) | **0.780** (0.744-0.817) |
| Additive Hazard + Time Embedding | **0.865** (0.836-0.915) | **0.817** (0.785-0.87) | **0.802** (0.763-0.846) | 0.779 (0.74-0.82) | 0.778 (0.737-0.819) |

Finally, we plotted the ROC curves for 1 to 5-year AUC results in Fig. 9 comparing the baseline model, all models with the time-decay attention blocks, and RADMIL plus laterality. We have also included the results of the final TRINet model that combines all modifications, the risk progression model (i.e., additive hazard layer with time embedding), and the Mirai model (Yala et al., 2021).

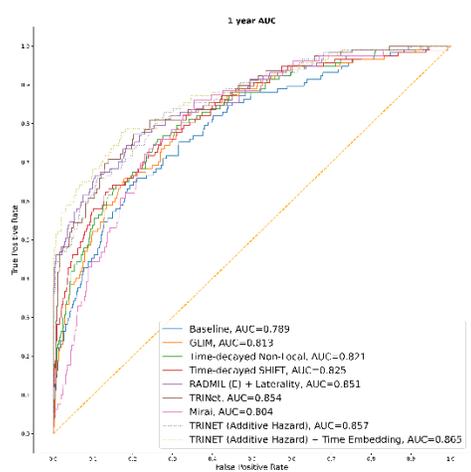
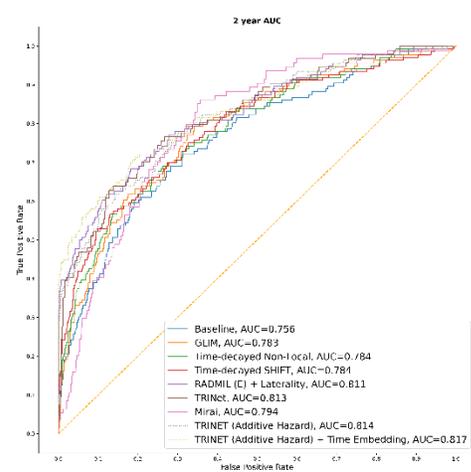
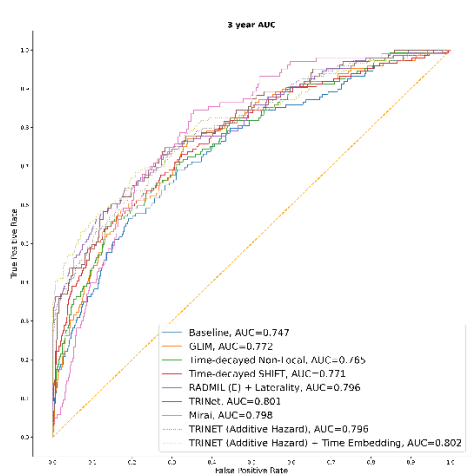
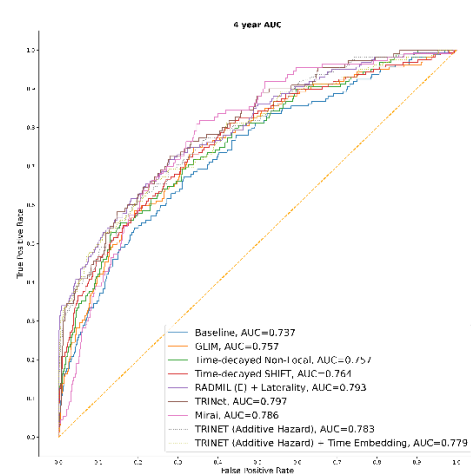
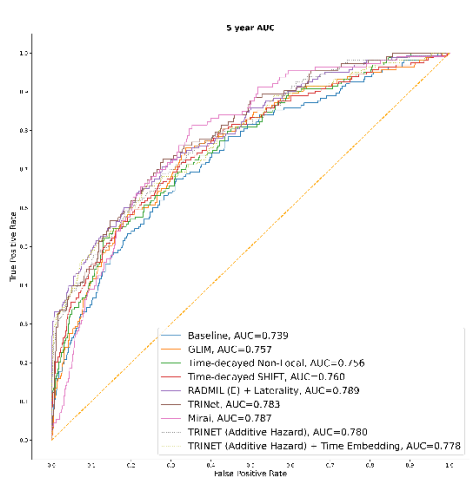

Fig. 9: ROC curves of ablation study results corresponding to 1 to 5-year AUC categories

## 4.5 CNN encoder ablation study

The results in Table 11 tabulate the ablation study on various baseline CNN encoders' performance in predicting cancer risk. We performed the ablation study on the most popular CNN models, namely AlexNet, VGG16, and ResNet18. We have denoted the models with all proposed modifications with asterisks in Table 11.

From Table 11, we observe increases across all five AUCs demonstrating that our proposed modifications using only the EMBED dataset significantly enhance the predictive power of all models. Resnet18 performs best among all encoders. This is not surprising as, notably, the state-of-the-art risk prediction method, Mirai (Yala et al., 2021) also utilizes ResNet18 as its CNN encoder, leveraging its strong feature extraction capabilities and residual connections to enhance risk prediction.

Table 11: AUC results with 95% CIs of different baseline CNN encoders used in our risk model.

|  | 1-year AUC | 2-year AUC | 3-year AUC | 4-year AUC | 5-year AUC |
|---|---|---|---|---|---|
| AlexNet | 0.691 (0.638-0.746) | 0.674 (0.623-0.728) | 0.678 (0.626-0.725) | 0.671 (0.623-0.717) | 0.675 (0.628-0.723) |
| AlexNet* | 0.741 (0.697-0.786) | 0.709 (0.663-0.757) | 0.708 (0.663-0.75) | 0.695 (0.649-0.739) | 0.697 (0.654-0.743) |
| VGG16 | 0.706 (0.663-0.754) | 0.689 (0.642-0.735) | 0.687 (0.647-0.729) | 0.676 (0.633-0.716) | 0.680 (0.638-0.719) |
| VGG16* | 0.753 (0.709-0.798) | 0.739 (0.697-0.781) | 0.736 (0.695-0.776) | 0.725 (0.688-0.763) | 0.724 (0.687-0.762) |
| Resnet18 | 0.789 (0.746-0.835) | 0.756 (0.713-0.799) | 0.747 (0.704-0.787) | 0.737 (0.696-0.780) | 0.739 (0.697-0.777) |
| Resnet18* | **0.851** (0.806-0.904) | **0.811** (0.764-0.862) | **0.796** (0.751-0.839) | **0.793** (0.751-0.84) | **0.789** (0.743-0.831) |

* Modifications added

## 4.6 Comparisons with State-Of-The-Art (SOTA) methods

We reimplemented the SOTA Mirai method (Yala et al., 2021) and evaluated it on the same dataset split as our study. Specifically, the test set from our dataset, as described in Section 3.1.1, was used for both our method and the reimplemented SOTA method for direct comparisons. The reported results from (Donnelly et al., 2024) are of a different dataset split, and may include different cohort selection, preprocessing pipelines and sample distributions. While we include these results for reference, they are not directly comparable to the results in this study.

Table 12: Comparison of AUC results with 95% CIs for our proposed method, the reimplemented SOTA method evaluated on the same dataset split as our method, and the reported results for the SOTA method from the original study. While the reimplementation ensures a fair comparison, the reported results from the original study are based on a different dataset split, which may involve variations in preprocessing, cohort selection, or test set composition.

| Method | 1-year AUC | 2-year AUC | 3-year AUC | 4-year AUC | 5-year AUC |
|---|---|---|---|---|---|
| Mirai (Donnelly et al., 2024) (Reported) | 0.84 (0.79-0.89) | 0.74 (0.70-0.78) | 0.72 (0.69-0.76) | 0.72 (0.69-0.75) | 0.71 (0.68-0.74) |
| AssymMirai (Donnelly et al., 2024) (Reported) | 0.79 (0.73-0.85) | 0.69 (0.65-0.73) | 0.68 (0.65-0.71) | 0.67 (0.64-0.70) | 0.66 (0.63-0.69) |
| Mirai (Yala et al., 2021) (Reimplemented) | 0.804 (0.761-0.852) | 0.794 (0.756-0.837) | 0.798 (0.759-0.838) | 0.786 (0.749-0.827) | 0.787 (0.746-0.824) |
| TRINet | **0.8549** (0.815-0.904) | **0.8139** (0.769-0.863) | **0.8014** (0.759-0.851) | **0.7971** (0.754-0.841) | **0.7934** (0.752-0.838) |

## 5. Discussion and Conclusions

This study improves upon our previous work on predicting cancer risk through sequential mammographic screenings (Yeoh et al., 2023). We present a number of new contributions in this study. First, we presented new advanced attention mechanisms that focus on time-decay (TD) attention. We propose new TD attention mechanisms for NL and SHIFT called TD-NL and TD-SHIFT, respectively. The TD attention prioritizes information from more recent screenings when extracting features from sequential mammograms as compared to previous screenings. Our findings highlight the critical role that incorporating time-sensitive information considerably enhances the effectiveness of NL and SHIFT attention mechanisms. Models equipped with TD attention consistently outperform their vanilla counterparts, demonstrating that embedding temporal information strengthens the model's ability to identify patterns relevant to cancer risk across various time intervals.

Second, our work also highlights the potential of the new RADMIL method for integrating radiomic features, proving essential for model prediction enhancement. This method improves upon the conventional method of integrating additional tabular input features through the FC layer only. Using RADMIL, the relative importance of each view can be weighed and aggregated appropriately. Conventionally, the CC view is more informative for computerized methods, whereas MLO is more useful for radiologists; thus different views provide differing information (Arefan et al., 2019; Mohamed et al., 2017; Tan et al., 2015). RADMIL helps the network to focus on relevant views to more accurately predict risk. Instead of focusing on one particular view, the model can decide on its own, through backpropagation training, useful views that provide relevant information for risk prediction.

Sharing a similar theme of directing the model's focus, we also incorporated lateral attention into RADMIL to direct the model to focus on which side of the breast would potentially develop cancer over time. While the uniform attention in AMIL allocates equal attentions to each view based on their overall contribution to cancer risk, lateral attention, trained through supervised learning, is fine-tuned to detect and highlight signs of cancer development in one breast side versus the other.

Fourth, understanding how risk evolves over time is critical for effective long-term monitoring and early intervention of cancer risk prediction. To enhance the additive hazard layer, we introduced a novel modification: a time embedding layer within the hazard model. This time embedding layer adds a temporal sensitivity, which allows the model to account for the timing of previous screenings when predicting future risk. Our approach optimizes the risk scoring mechanism by embedding temporal information directly into the risk model enabling the additive hazard layer to adjust risk assessments in a nuanced manner based on screening intervals.

Fifth, we explored a label assignment strategy rooted in bilateral asymmetry detection, aiming to enhance the model's continual learning process by harnessing asymmetrical features between left and right breast views. This method adaptively refines its predictive accuracy through label reassignment based on laterality—allowing it to better capture risk factors that manifest as asymmetrical differences between breast sides.

Sixth, an ablation study on different CNN encoder architectures confirms the advantages of the modified ResNet18 over other popular alternatives including AlexNet and VGG16. The standout performance of ResNet18 reinforces the importance of using solid feature extractors in breast cancer risk models, as even small architectural improvements can significantly boost predictive capability. Given ResNet18's success in our study and in our state-of-the-art competitor, Mirai (Yala et al., 2021), it is a strong baseline encoder that can be used for future model iterations.

In conclusion, we developed a new temporally sensitive model for cancer risk prediction that incorporates time-decay attention mechanisms and integrates radiomic features through a novel architecture that uses attention-based multiple instance learning called RADMIL. Furthermore, our model's innovative use of a time-embedded additive hazard layer, and a new self-training method using asymmetry-based label assignment for continual learning contributes valuable frameworks for improving risk assessment based methods in the literature. A downside of our model is that we were only able to test it on two public datasets of American (EMBED) and Swedish (CSAW) populations, which may limit generalizability, highlighting the need for broader clinical testing and validation on more diverse patient datasets in the future. However, the results still demonstrate that our new model incorporating these innovative methods improves cancer risk assessment, outperforms SOTA methods, and promotes new approaches for personalized medicine and/or screening regimens in individual women.

## Acknowledgement

This work was supported by the Fundamental Research Grant Scheme (FRGS), Ministry of Higher Education Malaysia (MOHE), under grant FRGS/1/2022/ICT02/MUSM/02/1.

## References

Anandarajah, A., Chen, Y., Colditz, G.A., Hardi, A., Stoll, C.R.T., Jiang, S., 2021. Studies of parenchymal texture added to mammographic breast density and risk of breast cancer: a systematic review of the methods used in the literature. Breast Cancer Research : BCR 24.

Arasu, V.A., Habel, L.A., Achacoso, N.S., Buist, D.S.M., Cord, J.B., Esserman, L.J., Hylton, N.M., Glymour, M.M., Kornak, J., Kushi, L.H., Lewis, D.A., Liu, V.X., Lydon, C.M., Miglioretti, D.L., Navarro, D.A., Pu, A., Shen, L., Sieh, W., Yoon, H.-C., Lee, C., 2023. Comparison of Mammography AI Algorithms with a Clinical Risk Model for 5-year Breast Cancer Risk Prediction: An Observational Study. Radiology 307 5, e222733.

Arefan, D., Mohamed, A.A., Berg, W.A., Zuley, M.L., Sumkin, J.H., Wu, S., 2019. Deep learning modeling using normal mammograms for predicting breast cancer risk. Medical physics.

Bahdanau, D., Cho, K., Bengio, Y., 2014. Neural Machine Translation by Jointly Learning to Align and Translate. CoRR abs/1409.0473.

Bond, M., Pavey, T.G., Welch, K., Cooper, C.E., Garside, R., Dean, S.G., Hyde, C., 2013. Systematic review of the psychological consequences of false-positive screening mammograms. Health technology assessment 17 13, 1-170, v-vi.

Carneiro, G., Nascimento, J.C., Bradley, A.P., 2015. Unregistered Multiview Mammogram Analysis with Pre-trained Deep Learning Models, International Conference on Medical Image Computing and Computer-Assisted Intervention.

Dadsetan, S., Arefan, D., Berg, W.A., Zuley, M.L., Sumkin, J.H., Wu, S., 2022. Deep learning of longitudinal mammogram examinations for breast cancer risk prediction. Pattern Recognit. 132, 108919.

Dembrower, K., Lindholm, P., Strand, F., 2020. A Multi-million Mammography Image Dataset and Population-Based Screening Cohort for the Training and Evaluation of Deep Neural Networks-the Cohort of Screen-Aged Women (CSAW). J Digit Imaging 33, 408-413.

Ding, Z., Dong, H., 2020a. Challenges of Reinforcement Learning.

Ding, Z., Dong, H., 2020b. Challenges of Reinforcement Learning, In: Dong, H., Ding, Z., Zhang, S. (Eds.), Deep Reinforcement Learning: Fundamentals, Research and Applications. Springer Singapore, Singapore, pp. 249-272.

Donnelly, J., Moffett, L., Barnett, A.J., Trivedi, H., Schwartz, F.R., Lo, J.Y., Rudin, C., 2024. AsymMirai: Interpretable Mammography-based Deep Learning Model for 1-5-year Breast Cancer Risk Prediction. Radiology 310 3, e232780.

Dosovitskiy, A., Beyer, L., Kolesnikov, A., Weissenborn, D., Zhai, X., Unterthiner, T., Dehghani, M., Minderer, M., Heigold, G., Gelly, S., Uszkoreit, J., Houlsby, N., 2021. An Image is Worth 16x16 Words: Transformers for Image Recognition at Scale. International Conference on Learning Representations.

Gastounioti, A., Conant, E.F., Kontos, D., 2016. Beyond breast density: a review on the advancing role of parenchymal texture analysis in breast cancer risk assessment. Breast Cancer Res 18, 91.

Gulcehre, C., Paine, T.L., Srinivasan, S., Konyushkova, K., Weerts, L., Sharma, A., Siddhant, A., Ahern, A., Wang, M., Gu, C., Macherey, W., Doucet, A., Firat, O., Freitas, N.d., 2023. Reinforced Self-Training (ReST) for Language Modeling. ArXiv abs/2308.08998.

Habib, A.R., Grady, D., Redberg, R.F., 2021. Recommendations From Breast Cancer Centers for Frequent Screening Mammography in Younger Women May Do More Harm Than Good. JAMA internal medicine.

Hayward, J.H., Ray, K.M., Wisner, D.J., Kornak, J., Lin, W., Joe, B.N., Sickles, E.A., 2016. Improving Screening Mammography Outcomes Through Comparison With Multiple Prior Mammograms. AJR. American journal of roentgenology 207 4, 918-924.

Hu, X., Zhang, L.-X., Gao, L., Dai, W., Han, X., Lai, Y.-K., Chen, Y., 2023. GLIM-Net: Chronic Glaucoma Forecast Transformer for Irregularly Sampled Sequential Fundus Images. IEEE Transactions on Medical Imaging 42, 1875-1884.

Huang, Z., Wessler, B.S., Hughes, M.C., 2023. Detecting Heart Disease from Multi-View Ultrasound Images via Supervised Attention Multiple Instance Learning, In: Kaivalya, D., Madalina, F., Shalmali, J., Zachary, L., Rajesh, R., Iñigo, U., Serene, Y. (Eds.), Proceedings of the 8th Machine Learning for Healthcare Conference. PMLR, Proceedings of Machine Learning Research, pp. 285--307.

Ilse, M., Tomczak, J.M., Welling, M., 2018. Attention-based Deep Multiple Instance Learning, International Conference on Machine Learning.

Jeong, J.J., Vey, B.L., Bhimireddy, A.R., Kim, T., Santos, T., Correa, R., Dutt, R., Mosunjac, M., Oprea-Ilies, G., Smith, G., Woo, M., McAdams, C.R., Newell, M.S., Banerjee, I., Gichoya, J.W., Trivedi, H., 2023. The EMory BrEast imaging Dataset (EMBED): A Racially Diverse, Granular Dataset of 3.4 Million Screening and Diagnostic Mammographic Images. Radiology. Artificial intelligence 5 1, e220047.

Keller, B.M., Nathan, D.L., Wang, Y., Zheng, Y., Gee, J.C., Conant, E.F., Kontos, D., 2012. Estimation of breast percent density in raw and processed full field digital mammography images via adaptive fuzzy c-means clustering and support vector machine segmentation. Med Phys 39, 4903-4917.

Kingma, D.P., Ba, J., 2015. Adam: A Method for Stochastic Optimization. CoRR abs/1412.6980.

Kontos, D., Winham, S.J., Oustimov, A., Pantalone, L., Hsieh, M.-K., Gastounioti, A., Whaley, D.H., Hruska, C.B., Kerlikowske, K., Brandt, K., Conant, E.F., Vachon, C.M., 2019. Radiomic Phenotypes of Mammographic Parenchymal Complexity: Toward Augmenting Breast Density in Breast Cancer Risk Assessment. Radiology 290 1, 41-49.

Lambin, P., Rios-Velazquez, E., Leijenaar, R.T.H., Carvalho, S., van Stiphout, R.G., Granton, P.V., Zegers, C.M.L., Gillies, R.J., Boellard, R., Dekker, A., Aerts, H.J.W.L., 2012. Radiomics: extracting more information from medical images using advanced feature analysis. European journal of cancer 48 4, 441-446.
Lee, H., Kim, J., Park, E., Kim, M., Kim, T., Kooi, T., 2023a. Enhancing Breast Cancer Risk Prediction by Incorporating Prior Images. ArXiv abs/2303.15699.
Lee, H., Kim, J., Park, E., Kim, M., Kim, T., Kooi, T., 2023b. Enhancing Breast Cancer Risk Prediction by Incorporating Prior Images. Springer Nature Switzerland, Cham, pp. 389-398.
Li, H., Robinson, K.R., Lan, L., Baughan, N.M., Chan, C.-W., Embury, M.D., Whitman, G.J., El-Zein, R., Bedrosian, I., Giger, M.L., 2023. Temporal Machine Learning Analysis of Prior Mammograms for Breast Cancer Risk Prediction. Cancers 15.
Li, T.Z., Xu, K., Gao, R., Tang, Y., Lasko, T.A., Maldonado, F., Sandler, K.L., Landman, B.A., 2022. Time-distance vision transformers in lung cancer diagnosis from longitudinal computed tomography, Medical Imaging.
Liu, J., Zhang, Y., Wang, K., Yavuz, M.C., Chen, X., Yuan, Y., Li, H., Yang, Y., Yuille, A.L., Tang, Y., Zhou, Z., 2024. Universal and Extensible Language-Vision Models for Organ Segmentation and Tumor Detection from Abdominal Computed Tomography. Medical image analysis 97, 103226.
Lotter, W., Diab, A.R., Haslam, B., Kim, J.G., Grisot, G., Wu, E., Wu, K., Onieva, J.O., Boyer, Y., Boxerman, J.L., Wang, M., Bandler, M., Vijayaraghavan, G., Gregory Sorensen, A., 2021. Robust breast cancer detection in mammography and digital breast tomosynthesis using an annotation-efficient deep learning approach. Nature Medicine 27, 244 - 249.
Ma, W., Zhao, Y., Ji, Y., Guo, X., Jian, X., Liu, P., Wu, S., 2019. Breast Cancer Molecular Subtype Prediction by Mammographic Radiomic Features. Academic radiology 26 2, 196-201.
Maghsoudi, O.H., Gastounioti, A., Scott, C., Pantalone, L., Wu, F.-F., Cohen, E.A., Winham, S.J., Conant, E.F., Vachon, C.M., Kontos, D., 2021. Deep-LIBRA: Artificial intelligence method for robust quantification of breast density with independent validation in breast cancer risk assessment. Medical image analysis 73, 102138.
McKinney, S.M., Sieniek, M., Godbole, V., Godwin, J., Antropova, N., Ashrafian, H., Back, T., Chesus, M., Corrado, G.C., Darzi, A., Etemadi, M., Garcia-Vicente, F., Gilbert, F.J., Halling-Brown, M.D., Hassabis, D., Jansen, S., Karthikesalingam, A., Kelly, C.J., King, D., Ledsam, J.R., Melnick, D.S., Mostofi, H., Peng, L.H., Reicher, J.J., Romera-Paredes, B., Sidebottom, R., Suleyman, M., Tse, D., Young, K.C., Fauw, J.D., Shetty, S., 2020. International evaluation of an AI system for breast cancer screening. Nature 577, 89 - 94.
Mirza, M., Osindero, S., 2014. Conditional Generative Adversarial Nets. ArXiv abs/1411.1784.
Mohamed, A.A., Luo, Y., Peng, H., Jankowitz, R.C., Wu, S., 2017. Understanding Clinical Mammographic Breast Density Assessment: a Deep Learning Perspective. Journal of Digital Imaging 31, 387-392.
Peng, B., Alcaide, E., Anthony, Q.G., Albalak, A., Arcadinho, S., Biderman, S., Cao, H., Cheng, X., Chung, M., Grella, M., Kranthikiran, G., He, X., Hou, H., Kazienko, P., Kocoń, J., Kong, J., Koptyra, B., Lau, H., Mantri, K.S.I., Mom, F., Saito, A., Tang, X., Wang, B., Wind, J.S., Wozniak, S., Zhang, R., Zhang, Z., Zhao, Q., Zhou, P., Zhu, J., Zhu, R., 2023. RWKV: Reinventing RNNs for the Transformer Era, Conference on Empirical Methods in Natural Language Processing.
Ren, W., Chen, M., Qiao, Y., Zhao, F., 2022. Global guidelines for breast cancer screening: A systematic review. The Breast : Official Journal of the European Society of Mastology 64, 85 - 99.
Ruder, S., 2016. An overview of gradient descent optimization algorithms. ArXiv abs/1609.04747.
Salz, T., Richman, A.R., Brewer, N.T., 2010. Meta‐analyses of the effect of false‐positive mammograms on generic and specific psychosocial outcomes. Psycho‐Oncology 19.
Shumailov, I., Shumaylov, Z., Zhao, Y., Gal, Y., Papernot, N., Anderson, R., 2023. The Curse of Recursion: Training on Generated Data Makes Models Forget. ArXiv abs/2305.17493.
Singh, A., Co-Reyes, J.D., Agarwal, R., Anand, A., Patil, P., Liu, P.J., Harrison, J., Lee, J., Xu, K., Parisi, A.T., Kumar, A., Alemi, A., Rizkowsky, A., Nova, A., Adlam, B., Bohnet, B., Sedghi, H., Mordatch, I., Simpson, I., Gur, I., Snoek, J., Pennington, J., Hron, J., Kenealy, K., Swersky, K., Mahajan, K., Culp, L., Xiao, L., Bileschi, M., Constant, N., Novak, R., Liu, R., Warkentin, T.B., Qian, Y., Dyer, E., Neyshabur, B., Sohl-Dickstein, J.N., Fiedel, N., 2023. Beyond Human Data: Scaling Self-Training for Problem-Solving with Language Models. Trans. Mach. Learn. Res. 2024.
Spathis, D., Kawsar, F., 2024. The first step is the hardest: pitfalls of representing and tokenizing temporal data for large language models. Journal of the American Medical Informatics Association.
Sun, Y., Dong, L., Huang, S., Ma, S., Xia, Y., Xue, J., Wang, J., Wei, F., 2023. Retentive Network: A Successor to Transformer for Large Language Models. ArXiv abs/2307.08621.

Tan, M., Mariapun, S., Yip, C.H., Ng, K.H., Teo, S.-H., 2019. A novel method of determining breast cancer risk using parenchymal textural analysis of mammography images on an Asian cohort. Physics in Medicine & Biology 64.

Tan, M., Pu, J., Cheng, S., Liu, H., Zheng, B., 2015. Assessment of a Four-View Mammographic Image Feature Based Fusion Model to Predict Near-Term Breast Cancer Risk. Annals of Biomedical Engineering 43, 2416-2428.

Tan, M., Zheng, B., Leader, J.K., Gur, D., 2016. Association Between Changes in Mammographic Image Features and Risk for Near-Term Breast Cancer Development. IEEE Transactions on Medical Imaging 35, 1719-1728.

Vaswani, A., Shazeer, N.M., Parmar, N., Uszkoreit, J., Jones, L., Gomez, A.N., Kaiser, L., Polosukhin, I., 2017. Attention is All you Need, Neural Information Processing Systems.

Wang, L., Zhang, X., Su, H., Zhu, J., 2023. A Comprehensive Survey of Continual Learning: Theory, Method and Application. IEEE Transactions on Pattern Analysis and Machine Intelligence 46, 5362-5383.

Wang, X., Girshick, R.B., Gupta, A., He, K., 2018. Non-local Neural Networks. 2018 IEEE/CVF Conference on Computer Vision and Pattern Recognition, 7794-7803.

Wu, C., Wu, F., Qi, T., Huang, Y., 2021. Fastformer: Additive Attention is All You Need.

Wu, N., Phang, J., Park, J., Shen, Y., Huang, Z., Zorin, M., Jastrzebski, S., Févry, T., Katsnelson, J., Kim, E., Wolfson, S., Parikh, U., Gaddam, S., Lin, L., Ho, K., Weinstein, J.D., Reig, B., Gao, Y., Toth, H., Pysarenko, K., Lewin, A., Lee, J., Airola, K., Mema, E., Chung, S., Hwang, E., Samreen, N., Kim, S., Heacock, L., Moy, L., Cho, K., Geras, K.J., 2020. Deep Neural Networks Improve Radiologists Performance in Breast Cancer Screening. IEEE transactions on medical imaging 39, 1184 - 1194.

Xie, Q., Hovy, E., Luong, M.-T., Le, Q.V., 2020. Self-Training With Noisy Student Improves ImageNet Classification. 2020 IEEE/CVF Conference on Computer Vision and Pattern Recognition (CVPR), 10684-10695.

Yala, A., Mikhael, P.G., Lehman, C.D., Lin, G., Strand, F., Wan, Y.-L., Hughes, K.S., Satuluru, S., Kim, T., Banerjee, I., Gichoya, J.W., Trivedi, H., Barzilay, R., 2022. Optimizing risk-based breast cancer screening policies with reinforcement learning. Nature medicine.

Yala, A., Mikhael, P.G., Strand, F., Lin, G., Smith, K., Wan, Y., Lamb, L., Hughes, K., Lehman, C., Barzilay, R., 2021. Toward robust mammography-based models for breast cancer risk. Science Translational Medicine 13.

Yeoh, H.H., Liew, A., Phan, R., Strand, F., Rahmat, K., Nguyen, T.L., Hopper, J.L., Tan, M., 2023. RADIFUSION: A multi-radiomics deep learning based breast cancer risk prediction model using sequential mammographic images with image attention and bilateral asymmetry refinement. ArXiv abs/2304.00257.

Yu, L., Simig, D., Flaherty, C., Aghajanyan, A., Zettlemoyer, L., Lewis, M., 2023. MEGABYTE: Predicting Million-byte Sequences with Multiscale Transformers. ArXiv abs/2305.07185.

Zhai, S., Talbott, W.A., Srivastava, N., Huang, C., Goh, H., Zhang, R., Susskind, J.M., 2021. An Attention Free Transformer. ArXiv abs/2105.14103.

Zheng, B., Sumkin, J.H., Zuley, M.L., Wang, X., Klym, A., Gur, D., 2012. Bilateral mammographic density asymmetry and breast cancer risk: a preliminary assessment. European journal of radiology 81 11, 3222-3228.

Zhu, X., Wolfgruber, T.K., Leong, L.T., Jensen, M., Scott, C.G., Winham, S.J., Sadowski, P., Vachon, C.M., Kerlikowske, K., Shepherd, J.A., 2021. Deep Learning Predicts Interval and Screening-detected Cancer from Screening Mammograms: A Case-Case-Control Study in 6369 Women. Radiology, 203758.